\documentclass[fleqn,usenatbib]{mnras}

\usepackage[T1]{fontenc}
\usepackage{ae,aecompl}

\usepackage{graphicx}	
\usepackage{amssymb}	

\newcommand\mearth{{\,{\rm M}_{\oplus}}}
\newcommand\mj{{\,{\rm M}_{\rm J}}}
\newcommand\simlt{\la}
\newcommand\simgt{\ga}
\newcommand\msun{M_\odot}

\title[Self-destruction of giant planets]{Tidal Downsizing Model. IV. Destructive feedback in planets.}

\author[Sergei Nayakshin]{
Sergei Nayakshin$^{1}$
\\
$^{1}$Department of Physics and Astronomy, University of
  Leicester, Leicester LE1 7RH, UK. {E-mail: sn85@le.ac.uk}
}

\date{Accepted XXX. Received YYY; in original form ZZZ}

\pubyear{2015}

\begin{document}
\label{firstpage}
\pagerange{\pageref{firstpage}--\pageref{lastpage}}
\maketitle

\begin{abstract}
I argue that feedback is as important to formation of planets as it is to
formation of stars and galaxies. Energy released by massive solid cores puffs
up pre-collapse gas giant planets, making them vulnerable to tidal disruptions
by their host stars.  I find that feedback is the ultimate reason for some of
the most robust properties of the observed exoplanet populations: the rarity
of gas giants at all separations from $\sim 0.1$ to $\sim 100$~AU, the
abundance of $\sim 10 \mearth$ cores but dearth of planets more massive than
$\sim 20 \mearth$. Feedback effects can also explain (i) rapid assembly of
massive cores at large separations as needed for Uranus, Neptune and the
suspected HL Tau planets; (ii) the small core in Jupiter yet large cores in
Uranus and Neptune; (iii) the existence of rare "metal monster" planets such
as CoRoT-20b, a gas giant made of heavy elements by up to $\sim 50$\%.
\end{abstract}


\section{Introduction}

From a single star to a cluster of galaxies, large self-gravitating
astrophysical systems are similar in two ways. Firstly, gravitational collapse
is the first step towards formation of these systems \citep[e.g., see][for
  formation of stars, galaxies and globular
  clusters]{Larson69,SZ72,WhiteRees78,FallRees85}. Secondly, most of these
systems suffer from energetic feedback released by very compact sub-parts of
the systems. For example, red giants are stars which experience a sudden
increase in the rate of nuclear reactions in or near the core. They swell by
up to $\sim 100$ times and eventually loose \citep[e.g.,][]{VW93} most of
their envelope, leaving behind very dense cores composed of heavy elements,
the white dwarfs. Negative feedback is also important during formation of
stars and star clusters
\citep[e.g.,][]{KrumholzEtal09,DaleandBonnell08}. Similarly, galaxies,
including our own, miss a significant amount of primordial gas believed to
have been expelled by energetic feedback \citep{CenOstriker99} due to stars
\citep[e.g.,][]{AgertzEtal13} and super-massive black holes
\citep{KingPounds15}.

Core Accretion (CA), the de-facto accepted theory for planet formation, \citep{Safronov72,PollackEtal96,IdaLin04b,MordasiniEtal09b}, excludes planets from this universal behaviour: planets grow bottom-up, and there is no scope for feedback. The first step in CA is assembly of  $\sim 1-100$ km sized rocky or icy bodies called planetesimals  \citep[for a recent review see, e.g.,][]{JohansenEtal14a}. These bodies coagulate to form planetary mass solid cores \citep{HayashiEtal85}, which then attract massive gas atmospheres from the surrounding disc. The atmospheres collapse and promote further gas accretion onto the massive cores \citep{Mizuno80,Stevenson82,HubickyjEtal05}. This eventually culminates in making gas giant planets. Massive cores are hence cradles of all types of planets in CA.

In this paper I argue for the opposite point of view in the context of the recently developed Tidal Downsizing (TD) theory for planet formation \citep{BoleyEtal10,Nayakshin10c}:
\begin{enumerate}
\item[(I)] All planets begin forming by gravitational collapse of gas-dominated fragments;
\item[(II)] Massive cores are the sources of energetic feedback, and are the main reason why so few gas giants exist.
\end{enumerate}

Point (I) has been discussed previously. In TD, Gravitational Instability (GI) of cold massive protoplanetary disc hatches gas fragments at distances of tens to hundreds of AU from the host star \citep{Kuiper51b,Boss97,Rice05,HelledEtal13a}. As fragments migrate towards the host \citep[e.g.,][]{BaruteauEtal11,MichaelEtal11,TsukamotoEtal14}, massive solid cores are assembled inside them by grain sedimentation \citep{HelledEtal08,Nayakshin10b}. Fragments that are disrupted by tidal forces from the host stars leave their rocky cores behind, which are, by large, fully made planets with mass from sub-Earth to many times larger \citep[e.g.,][]{BoleyEtal10,Nayakshin10b}. Fragments that managed to stay intact mature into jovian gas planets. 
Recent population synthesis calculations \citep{NayakshinFletcher15} found that many of the observed properties of exoplanet populations, previously claimed to be only accountable with CA, e.g., the positive correlation of gas giant planets with metallicity of the host stars \citep{FischerValenti05,IdaLin04b,MordasiniEtal09b},  over-abundance of metals in planets compared to the host stars  \citep[e.g.,][]{MillerFortney11}, planet mass function, etc., are also naturally explained by TD.

The main focus of this paper is point (II), that is,  the role played by feedback in TD. I perform numerical coupled protoplanetary disc evolution -- planet formation calculations very similar in set-up to those of \cite{NayakshinFletcher15}, but with three varying assumptions about core formation and their luminosity output. The "standard" simulation is nearly identical to those in the quoted paper but uses somewhat relaxed grain sedimentation assumptions, promoting faster core growth. The second simulation is identical to the first but the luminosity of the cores is set to be $10^5$ times lower, rendering core's feedback unimportant. Finally, in the simulation "no cores" core formation is turned off, which makes it similar to those in \cite{Nayakshin15b}, where no core formation was allowed, although with more disc physics included here.

I find that without core formation, or with core formation but with their energy release artificially suppressed, TD is unable to account for a number of observational facts, such as the observed planet and core mass functions, constraints from direct imaging surveys on the frequency of gas giants at large separations, existence of large scale debris discs, etc.

With the energy release by the core included in the models,  planet's evolution takes two divergent paths depending on the luminosity (mass) of the core. If core's mass is below $\sim$ a few $\mearth$, the core is unimportant for the fragment, and it is pretty much a passive passenger of the fragment. However, if and when the core's mass exceeds  $\sim 5 \mearth$, its energy release starts to affect the host gas fragment by slowing down its contraction and even reversing it into expansion. This leads to most of fragments with cores more massive than a Super Earth being tidally disrupted.

This result has a number of attractive observational implications:
\begin{enumerate}
\item[(1)] Super-Earth to Neptune mass planets are made in very frequent gas fragment disruptions, so they are the most common type of planets \citep{MayorEtal11,BatalhaEtal13}.

\item[(2)] The mass of $\sim 10-20 \mearth$ forms the barrier above which most solid cores cannot grow or else they destroy the parent fragment. The planet and core mass functions predicted by TD scenario both nose-dive above these masses, as observed \citep{MayorEtal11,HowardEtal12}.

\item[(3)] Massive gas giants in separations of $10-100$~AU are rare in direct imaging surveys \citep[e.g.,][]{BillerEtal13,BowlerEtal15}. This is usually interpreted as evidence that massive cold protoplanetary discs do not make gas fragments by GI frequently \citep[e.g.,][]{RiceEtal15}. However, I find here that with introduction of feedback many of the gas giants that would have been otherwise observable in such surveys get disrupted, leaving behind Super Earth to $\simlt$ Saturn mass planets, which are not directly observable yet. 

\item[(4)] The disruption remnants are produced rapidly (in under 1 Myr), which is a bonus since it may explain formation of planets with properties consistent with Uranus and Neptune in the Solar System and the suspected HL Tau planets. In \S \ref{sec:HLTAU} I detail arguments why TD scenario for making HL Tau appears to be much preferable to anything that can be offered by  CA or GI scenarios; 

\item[(5)] Although discussed in a separate paper (Fletcher \& Nayakshin 2015, to be submitted), same disruptions yield debris rings with radii from a few to over $ 100$ AU \citep[see also][]{NayakshinCha12}. These debris discs have metallicity correlation properties consistent with observations.

\item[(6)] The largest giant planet in the Solar System, Jupiter, appears to have the smallest core \citep{Guillot05}. This is peculiar in the context of CA but is natural in the view of results presented below. Neptune and Uranus cores grew too large, and hence their Hydrogen envelopes were expelled due to the intense feedback from their cores \citep[this was suggested already by][]{HW75}. Jupiter avoided this fate because its core is far less massive;

\item[(7)] It turns out that for a small number of fragments, pebble accretion onto the fragment and the core grain growth rate adjust to one another, and hence the fragment can accrete grains without being disrupted for a long time. This creates particularly metal-rich planets, with metallicities up to $Z_{\rm pl}\sim 0.4-0.5$. I propose this scenario for formation of the "metal monster" gas giant CoRoT-20b \citep{DeleuilEtal12}.
\end{enumerate}

I therefore suggest that negative feedback from massive cores during formation phase of planets is a key process that deserves a correspondingly important place in theory of planet formation.


Section \ref{sec:feedback} presents analytical arguments explaining why feedback from massive cores ought to be important for pre-collapse gas giant planets, and derives the critical core mass, $M_{\rm sd}$, at which a fragment would self-disrupt. In \S \ref{sec:rhd}, an isolated planet case, fed by pebbles at a high rate, is simulated with a radiative hydrodynamics approach, confirming the analytical suggestions. I then go on to discuss numerical methods to study the coupled planet-disc evolution in \S \ref{sec:methods} and show an example protoplanetary disc -- fragment calculation in which the fragment is disrupted due to feedback. \S \ref{sec:synthesis} introduces the population synthesis model that is used to study planet formation outcomes of the model in a statistical sense. In \S \ref{sec:results}, results of the simulations are discussed in detail. Finally, \S \ref{sec:discussion} considers observational implications of this paper's results.

\section{Analytical arguments}\label{sec:feedback}

Newly made GI fragments have virial temperatures in hundreds of Kelvin, when Hydrogen is molecular, and gas densities $\sim 10$ orders of magnitude lower than the density of present day Jupiter \citep{Bodenheimer74,HelledEtal08,Nayakshin10a}. Depending on fragment mass and dust opacity \citep[e.g.,][]{HB11}, it takes them $\sim 1$~Million years to contract and collapse by H$_2$ molecule dissociation  into the "hot start" stage usually taken as the beginning of a GI planet's life \citep[e.g.,][]{MarleyEtal07}.  The cool and extended stage of GI fragment evolution is logically called the pre-collapse stage. To become a hot jupiter, the fragment needs to collapse or else it is disrupted, typically at a few AU distance from the star \citep{Nayakshin10c}. 

The low temperature of pre-collapse fragments implies that very little energy is needed to unbind them: per unit mass, this energy is 4-6 orders of magnitude smaller than that needed to unbind a galaxy. The latter can only be affected significantly by energy produced by nuclear reactions in stars or by black hole accretion. We shall see now that accretion energy of a massive core is all that is needed to unbind pre-collapse gas giants.

The total energy of the fragment (pre-collapse planet) of mass $M_{\rm p}$
and radius $R_{\rm p}$ can be estimated as that of a polytropic sphere with index $n=5/2$:
\begin{equation}
E_{\rm tot} = - {3-n\over 5-n} {G M_{\rm p}^2 \over R_{\rm p}} \approx - 2\times 10^{40}
\; {\rm erg} \;T_3 \;\left({M_p\over 1 \mj}\right)\;,
\label{etot0}
\end{equation}
where $T_3 = T_c/(1000$~K) is the central temperature of the planet. 
The
gravitational potential energy of a dense core of mass $M_{\rm core}$,
modelled as a sphere of a uniform density $\rho_0$ (in g/cm$^3$) and 
radius $R_{\rm core}$, is
$E_{\rm core} \sim G
  M_{\rm core}^2/(2 R_{\rm core}) \approx 5 \times 10^{40} (M_{\rm core}/10
  \mearth)^{5/3} \rho_0^{1/3}$
erg. When the core cools, this energy is released and enters the gas envelope surrounding it (the fragment). If the fragment is unable to radiate this input energy quickly enough into the surrounding space, then it will be unbound provided  $E_{\rm core} > |E_{\rm
  tot}|$. This condition can be used to set the maximum ``self-disruption'' mass of the core:
\begin{equation}
M_{\rm sd} = 5.8 \mearth \left({T_3 M_{\rm p} \over 1 \mj}\right)^{3/5}
\rho_0^{-1/5}\;.
\label{Mcrit}
\end{equation}
 We shall see that this is a robust limit for our models, and that in practice cores can hardly exceed mass of $\sim20\mearth$. This results be anticipated as following. The fragment's central temperature, $T_{\rm c}$, cannot exceed $\sim 2,000$~Kelvin or the fragment would collapse through H$_2$ dissociation \citep[e.g.,][]{Bodenheimer74}; hence, $T_3\lesssim 2$. The core cannot grow by grain sedimentation in post-collapse fragments, of course, since even most refractive grain species vaporise above temperature $\sim 1500$~K \citep[see][for detailed description of grain physics used here]{Nayakshin14b}. Further, most massive cores are actually assembled in moderately massive fragments, $M_{\rm p}\lesssim$ a few Jupiter masses \citep[see fig. 18 in][]{NayakshinFletcher15} since fragments of higher masses contract more rapidly, becoming too hot for grain sedimentation, and making {\em smaller} cores than their less massive counterparts \citep[see ][]{HS08,Nayakshin10b}. For example, a fragment of $4 \mj$ contracts radiatively $\sim 10^3$ times more rapidly than a fragment of mass $0.5\mj$ \cite[e.g., see fig.1 in][]{Nayakshin15a}. Addition of pebble accretion accelerates this collapse even further. Therefore, the factor in brackets in equation \ref{Mcrit} does not exceed $\sim 6$ or so.
 
 Making parallels to other astrophysical objects, equation \ref{Mcrit} is akin the so-called M-$\sigma$ limit for super-massive black holes; black holes more massive than that expel  the gas from the parent galaxy \citep[e.g.,][]{SilkRees98,King03}. One may hence expect that the mass of $\sim 10\mearth$ will stand out as an important feature in the observed planet populations.
 
\section{Calculations}\label{sec:calculations}

I use two computational methods to check analytical predictions. In the first, an explicit 1D radiative hydrodynamics (RHD) code is used to model the planet only. The second (\S \ref{sec:methods}) simplifies the treatment of the planet by assuming a hydrostatic balance and a fully convective structure for the planet, but also includes a time-dependent model for the disc and the planet-disc interaction. The first approach is more accurate as far as the planet evolution is concerned but is unfortunately too slow due to its explicit nature, the second is less accurate for the planet's internal variables but is more numerically expedient and can be applied to study statistical outcomes of feedback for planet formation.

\subsection{A stand-alone planet calculation}\label{sec:rhd}

I start with the simplest numerical setup: an isolated planet of a fixed gas mass that accretes pebbles at a specified rate, $\dot M_{\rm z}$. The rate is prescribed via time scale $t_{\rm z}$:
\begin{equation}
\dot M_{\rm z}\equiv {Z_\odot M_{\rm p}(0)\over t_{\rm z}}\;,
\label{tz}
\end{equation}
where $Z_{\odot} = 0.015$ is Solar metallicity \citep{Lodders03}, and $M_{\rm p}(0)$ is the initial total mass of the planet. In this section only, I use the same planet evolution code as that described in \S 3 of \cite{Nayakshin15a}.  Spherically symmetric Lagrangian hydrodynamics in 1D approximation is used to model the evolution of both gas and four grain species (water, organics, rocks and Fe). Grains are allowed to grow by sticking collisions and can get fragmented in high speed collisions or vaporised if the surrounding gas is too hot \citep[see for more detail][]{Nayakshin14b}. The equation of state for the gas includes H$_2$ molecule rotational and vibrational transitions, dissociation, and H atom ionisation. Here we use interstellar grain opacity multiplied by a factor of $0.1$ to allow for grain growth, but the results are weakly dependent on opacity unless dust opacity is another order of magnitude smaller \citep[see][for a discussion of the role of grain opacity in TD]{Nayakshin15c}. 

The calculation presented here is identical to similar runs presented in \cite{Nayakshin15a}, except that core growth was turned off in the cited paper, while here it is allowed. While the gas fragment is modelled explicitly, the central core is simply set to have a fixed material density $\rho_0 = 3$ g/cm$^3$. The gas fragment has two boundary conditions, one at $R= R_{\rm core}$, the core radius, and the other at the outer radius of the planet, $R=R_{\rm p}$. On the inner boundary, the luminosity is set to $L(R_{\rm in})= L_{\rm core}$, the core luminosity (see \S \ref{S:cores}). The total energy of the fragment hence decreases due to radiative losses from the surface, $L_{\rm rad}$, but increases due to energy injection from the core at rate $L_{\rm core}$. 

Finally, pebbles accreting onto the fragment add additional mass and gravitational potential energy to the fragment \citep[see ][]{Nayakshin15a,Nayakshin15b}. The latter term is negative. The positive term due to the deposition of pebble's kinetic energy into the planet is very small because the pebble's velocity is not the free fall one as it would be if pebbles accreted onto the planet in vacuum, but is given by the sedimentation velocity in the extended atmosphere around the planet, and also cannot exceed grain breaking velocity of a few to a few tens of m/sec. 

To understand the fragment's evolution, consider the total energy conservation equation for it\footnote{This equation is not solved directly by the RHD code, but is obeyed due to energy conservation. This equation is actually used for evolving the fragment in the simpler "follow adiabats approach" utilised in planet-disc co-evolution experiments as detailed in \S \ref{sec:methods}.},
\begin{equation}
{d E_{\rm tot}\over dt} = - L_{\rm rad} + L_{\rm core} - L_{\rm peb} \;,
\label{etot1}
\end{equation}
where the last term is the fragment's potential energy change due to the pebble accretion onto the planet
at the rate $\dot M_z> 0$, which has dimensions of a luminosity: 
\begin{equation}
L_{\rm peb} = {G M_{\rm p}\dot M_z \over R_{\rm p}}\;.
\end{equation}
This term is negative because the potential energy gained by the fragment as pebbles are added is negative. For moderately
massive gas giants, $M_{\rm p} \simlt$ a few $\mj$, $L_{\rm rad}$ is small, and pebble accretion is usually the dominant effective cooling
mechanism for such fragments \citep{Nayakshin15a}.  Clearly, when the right hand side of equation \ref{etot1} is negative, the fragment contracts, and vice versa.

\begin{figure}
\includegraphics[width=0.99\columnwidth]{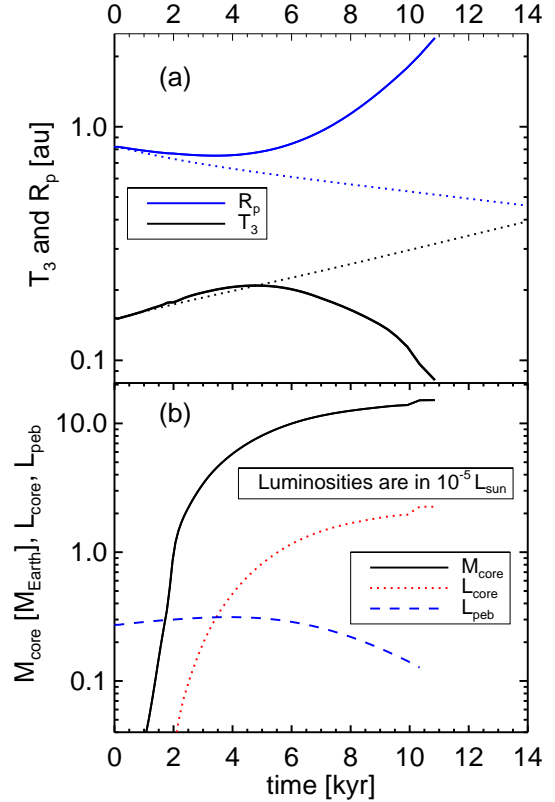}
 \caption{Panel (a) shows the gas fragment central temperature $T_3 = T_{\rm c}/10^3 K$, and planet radius, 
   $R_{\rm p}$, versus time for simulations with (solid curves) and without (dotted) core formation, as described in \S \ref{sec:rhd}. 
   Panel (b) shows the core mass, $M_{\rm core}$, core luminosity,
   $L_{\rm core}$, and pebble luminosity, $L_{\rm peb}$, for the simulation with core formation and feedback. Note that the fragment starts to expand when $L_{\rm core}$ exceeds $L_{\rm peb}$.}
   \label{fig:rhd}
 \end{figure}

In the experiments shown in this section, the initial cloud mass, metallicity and central temperature are $M_{\rm p}=1\mj$, $Z_\odot$ and $150$~K, respectively. The metal loading time scale is set to $t_{\rm z} = 2000$~years. Figure \ref{fig:rhd} compares two runs, one without grain growth and without core formation \citep[so identical in setup to][]{Nayakshin15a}, and the other with grain growth and core formation allowed. Panel (a) of the figure shows in black colour the evolution of $T_3$, the central fragment's temperature measured in $10^3$~K, and the planet's radius, $R_{\rm p}$ [au], shown with blue curves. The solid curves show the case of the fragment with the core, whereas the dotted ones correspond to the core-less fragment. Panel (b) of fig. \ref{fig:rhd} shows the core mass, in Earth masses, and the luminosity of the core and the pebble accretion luminosity, as indicated in the legend. The units for the luminosity curves are $10^{-5}L_\odot$. 

Core's growth initially proceeds very slowly because the initial size of grains in our calculations is $10^{-2}$~cm, and the grains first grow mainly due to Brownian motion of smaller grains sticking to the larger grains. However, once grains reach the size of $\sim 1$~cm, they are able to sediment much quicker and hence core's growth takes off at that point.

The comparison of the black and the blue curves in panel (a) shows that massive core formation reverses the fragment collapsing trend (that is, $R_{\rm p}$ decreasing while $T_{\rm c}$ increases) when the core's luminosity exceeds $L_{\rm peb}$. By the end of the calculation the fragment is completely unbound. The mass of the core reached by that point is $M_{\rm core} = 15.2 \mearth$, which is consistent to what is expected based on equation \ref{Mcrit}. The last episode of core growth before it gets disrupted in this simulation is due to a late burst of $1.3\mearth$ water ice accretion onto the core. This becomes possible when the central temperature drops low enough for water to condense out as ice. Such later water ice accretion events are never found in simulations without feedback because the fragments only heat up with time in such simulations and the fragments' central parts are always too hot for water ice \citep[this was first pointed out by][]{HelledEtal08}.

This simulation confirms that massive and luminous core formation may lead to disruption of the fragments as envisaged in the analytical argument. It is interesting to now consider the more complicated situation when pebble accretion is determined by the co-evolution of the planet and the disc. In such more complicated situation, the planet migrates in and may open a gap, while the protoplanetary disc looses mass to the star and to a photo-evaporative wind. The spectrum of outcomes is hence much richer than it is for the isolated (albeit accreting pebbles) planet that we considered in this section.

\subsection{Planet-disc evolution experiments}\label{sec:methods}

To calculate the evolution of both the protoplanetary disc and the planet, I use a 1D viscous disc plus a 1D spherically symmetric planet evolution codes coupled as described in \cite{Nayakshin15c} and in the appendix \ref{S:Appendix}. The planet evolution module is simplified in terms of gas fragment evolution compared to the method used in \S \ref{sec:rhd}, which is too slow and computationally expensive to combine with a protoplanetary disc evolution code. The method used here and in the rest of the paper is to assume that the fragment is strongly convective and therefore has a uniform specific entropy. The latter changes with time, of course, as the fragment contracts or expands.

In short, a simulation begins with a gas fragment born in the outer disc at $a\sim 100$ au.  Gravitational torques from the disc push the fragment in. I calculate pebble accretion rate onto the fragment, fragment's irradiation, and internal processes, such as grain growth, sedimentation and core assembly. This determines the rate of fragment contraction and its other properties. At the same time, the disc mass decreases due to accretion onto the star and photo-evaporation. The simulation stops when the disc is dissipated away or the fragment reaches the innermost radius of the computational domain (set at $R_{\rm in} = 0.1$ AU here). Pebble accretion onto the gas fragment is 
a particularly important addition to TD framework, due to which gas fragments become metal rich, make more massive cores, and collapse faster \citep{Nayakshin15a} than due to the usually considered radiative cooling alone \citep[e.g.,][]{Bodenheimer74}. 

As with the RHD code used in \S \ref{sec:rhd}, the planet evolution module of the code (see appendix \ref{S:planet}) deals with the gas-dust part of the planet explicitly but assumes the central core to have a fixed material density $\rho_0 = 3$ g/cm$^3$. The planet's properties are evolved by solving the energy conservation equation \ref{etot1}.

 Expansion of a fragment while it continues to migrate inwards tantamount to suicide since the Hill's radius of the planet, $R_{\rm H} = a (M_{\rm p}/3M_*)^{1/3}$,  keeps shrinking and the planet will be disrupted when it fills its Roche lobe. Since pebble accretion drives planet contraction, and since higher metallicity hosts provide higher $\dot M_z$, gas fragment survival against tidal disruption is much more likely at high metallicities. This explains \citep{Nayakshin15b,NayakshinFletcher15}, in the context of TD, why gas giants at small separations are found to be almost exclusively around metal rich hosts.


\begin{figure}
\includegraphics[width=0.8\columnwidth]{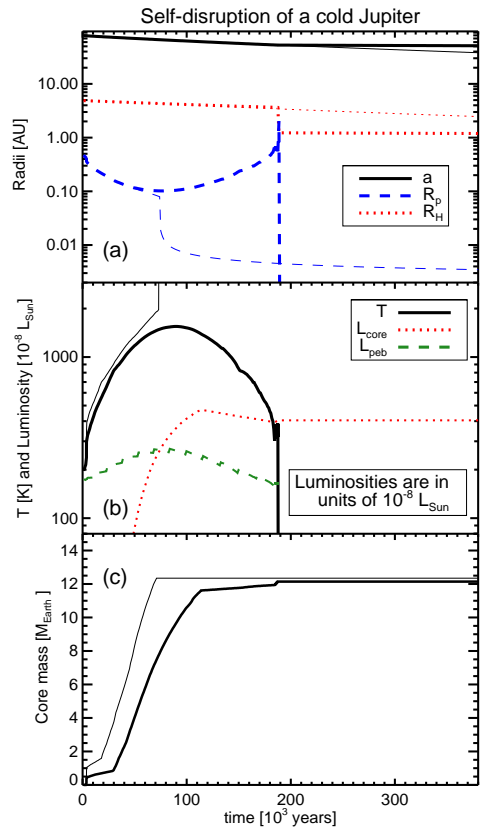}
 \caption{Thick curves: evolution of a fragment in one of the ST simulations
   that ends with a massive core disrupting the fragment at about 0.2
   Myr. Panel (a) shows the planet-star separation (black solid), $R_H$ and
   $R_p$, versus time; (b) shows the central temperature $T_c$, $L_{\rm core}$
   and $L_{\rm peb}$; (c) shows the core's mass. Thin curves are for an identical simulation but 
   with $L_{\rm core}$ arbitrarily reduced by a factor of 100. In the latter case the fragment manages to collapse and
   survives as a gas giant planet at $a\approx 10.5$~AU.}
   \label{fig2}
 \end{figure}

To exemplify the self-destructive fragment evolution anticipated in \S \ref{sec:feedback}, a single planet formation experiment with our defaults assumptions (picked from suite of simulations "ST", see below) is shown in fig. \ref{fig2} with thick curves. This calculation starts with a gas fragment born at $a\approx 80$~AU with initial mass of $M_{\rm p} \approx 0.7
\mj$. The fragment initially contracts due to pebble accretion, but when the core feedback becomes too powerful, the fragment self-inflates, gets disrupted, and leaves behind a massive ($M_{\rm core} \approx 12 \mearth$) core. This transformation of the fragment from a potential gas giant into a "just" a massive core occurs in only $\sim$ 0.2 Myr, although the core stops migrating only when the disc is dissipated away.

The black, the red and the blue curves in panel (a) show the time evolution of the planet-host separation, $a$, Hill ($R_{\rm H}$) 
and planet ($R_{\rm p}$) radii, respectively.  

The black curve in panel (b) shows the fragment's
central temperature, $T_{\rm c}$, whereas the red and the green show $L_{\rm core}$
and $L_{\rm peb}$, respectively. The kins and jumps in the $T_{\rm c}$ curve are due to grains vaporising or re-forming \citep[our equation of state takes into account the latent heat of grain vaporisation for the three grain species we consider; see][]{Nayakshin14b}. The radiative luminosity of the fragment is much smaller than the other two dominant terms ($L_{\rm peb}$ and $L_{\rm core}$) for this calculation. While $L_{\rm core} \ll L_{\rm
  peb}$, $T_{\rm c}$ keeps increasing with time and $R_{\rm p}$ decreases. This construction is due to pebble
accretion and is similar to the core-free evolution studied in \cite{Nayakshin15a}.  However, once $L_{\rm core} > L_{\rm peb}$, the fragment
starts to expand, and its $T_c$ decreases. Since $R_{\rm p}$ increases as $R_{\rm H}$ decreases with
time, the fragment is eventually disrupted at $t\approx 0.19$~Myrs. Having destroyed its gas
envelope, the core is nearly naked: the dense atmosphere near the core \citep[calculated as in][]{NayakshinEtal14a} weighs only $M_{\rm atm} \sim 0.3 \mearth$.
The end result is a core-dominated planet, somewhat similar to but a little less massive than Neptune, at $a=44$~AU.

Overall the planet's evolution is this coupled planet-disc calculation is very reminiscent of the isolated planet run presented in \S \ref{sec:rhd}, although the pebble accretion rate onto the fragment is lower, and it takes it longer to build a massive enough core to self-disrupt.

To demonstrate that the planet's disruption is induced by the core's high luminosity for this run, even though the mechanics of the final disruption is due to tides from the host star, this calculation was repeated but this time the core's
luminosity was arbitrarily multiplied by $0.01$. Thin lines in fig. \ref{fig2} show the result. The core's luminosity is now negligible in
the thermal evolution of the fragment. The fragment heats up due to pebble accretion, and collapses by H$_2$ dissociation to much higher
densities and temperatures at time $t \approx 0.07$ Myr. Its radius drops to a few times that of Jupiter (note the thin blue dashed curve in panel a), which is now always much smaller than $R_{\rm H}$. With the much dimmer core, the planet is never disrupted. Although not shown in
fig. 2, the fragment migrates in to $a\approx 10.5$~AU by $t = 2.8$~Myr, when the protoplanetary disc dissipates away, to
become a gas giant with $M_p = 0.8 \mj$ (having accreted $\sim 0.1 \mj$ of
pebbles). Interestingly, the core's mass is also around $12\mearth$ but this core is entirely engulfed by the massive gas envelope. The core does not grow much beyond that of the original calculation because once the temperature reaches $T\simgt 1500$ K in the centre, all species of grains are vaporised in the centre, and the core cannot grow further by grain sedimentation \citep{Nayakshin10b}.

\section{Population synthesis}\label{sec:synthesis}

It is useful to explore how massive core feedback affects planet formation outcome in a statistical sense. To reach that goal, population synthesis approach is used, in which simulations similar to the one described in \S \ref{sec:methods} are repeated to cover a large parameter space of possible initial conditions for the disc mass, photo-evaporation rate, fragment's initial location, mass, etc. (see appendix \ref{S:popsyn}).  Each of the simulations begins
with a gas fragment, with a mass randomly sampled between $0.33$ and 8 $\mj$,
born at $70 < a < 105$~AU in a massive protoplanetary disc.

\begin{figure} 
\includegraphics[width=\columnwidth]{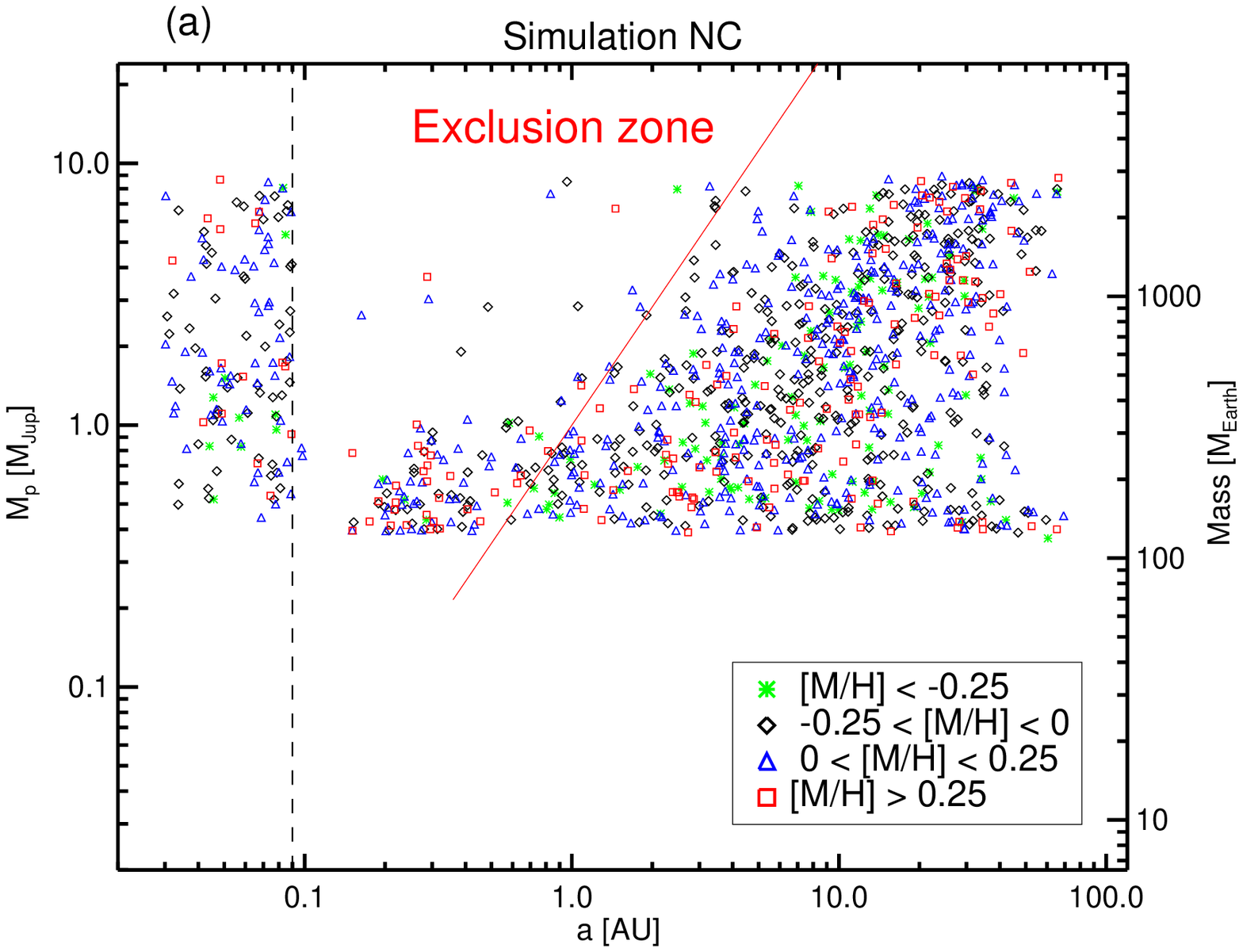}
\includegraphics[width=\columnwidth]{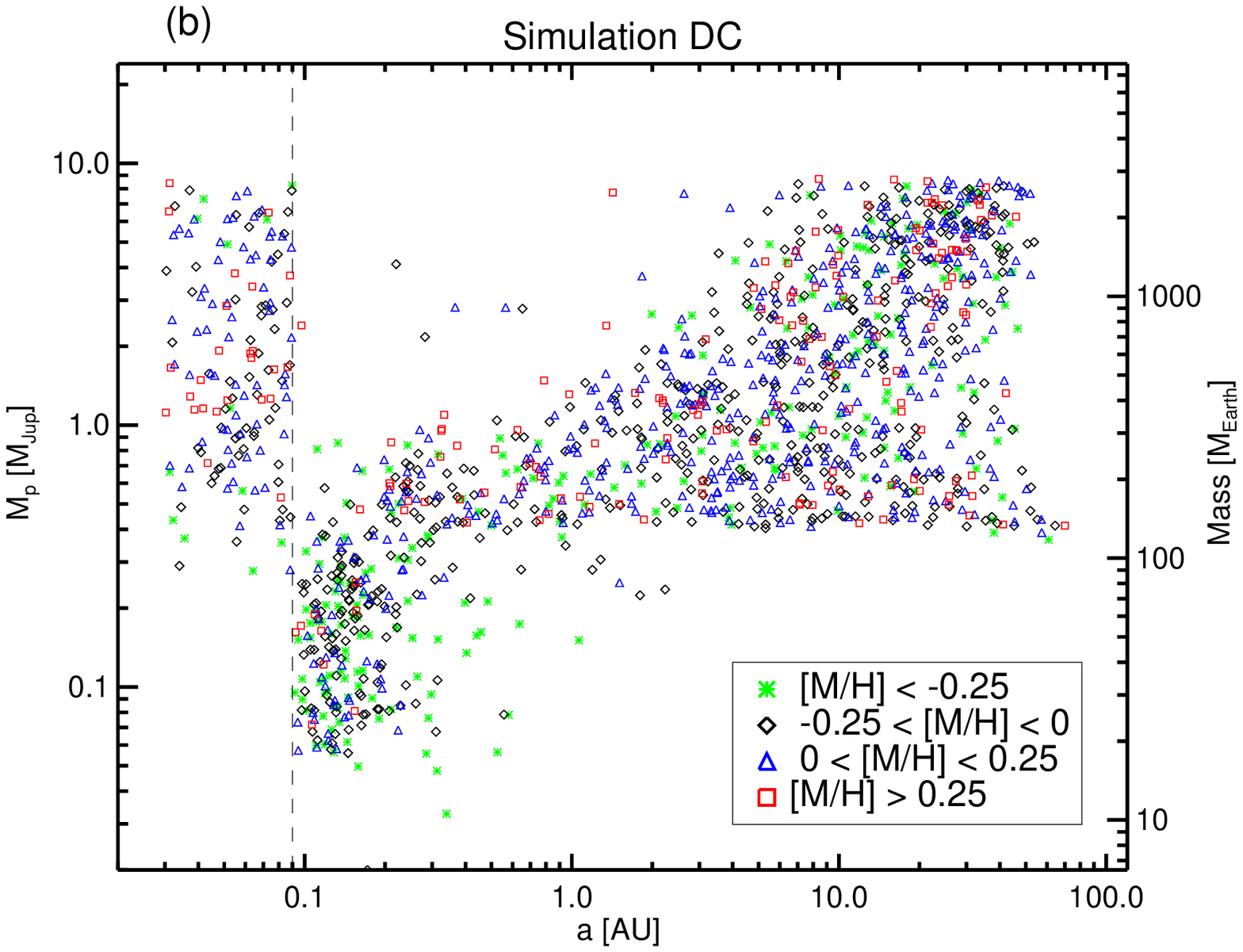}
\includegraphics[width=\columnwidth]{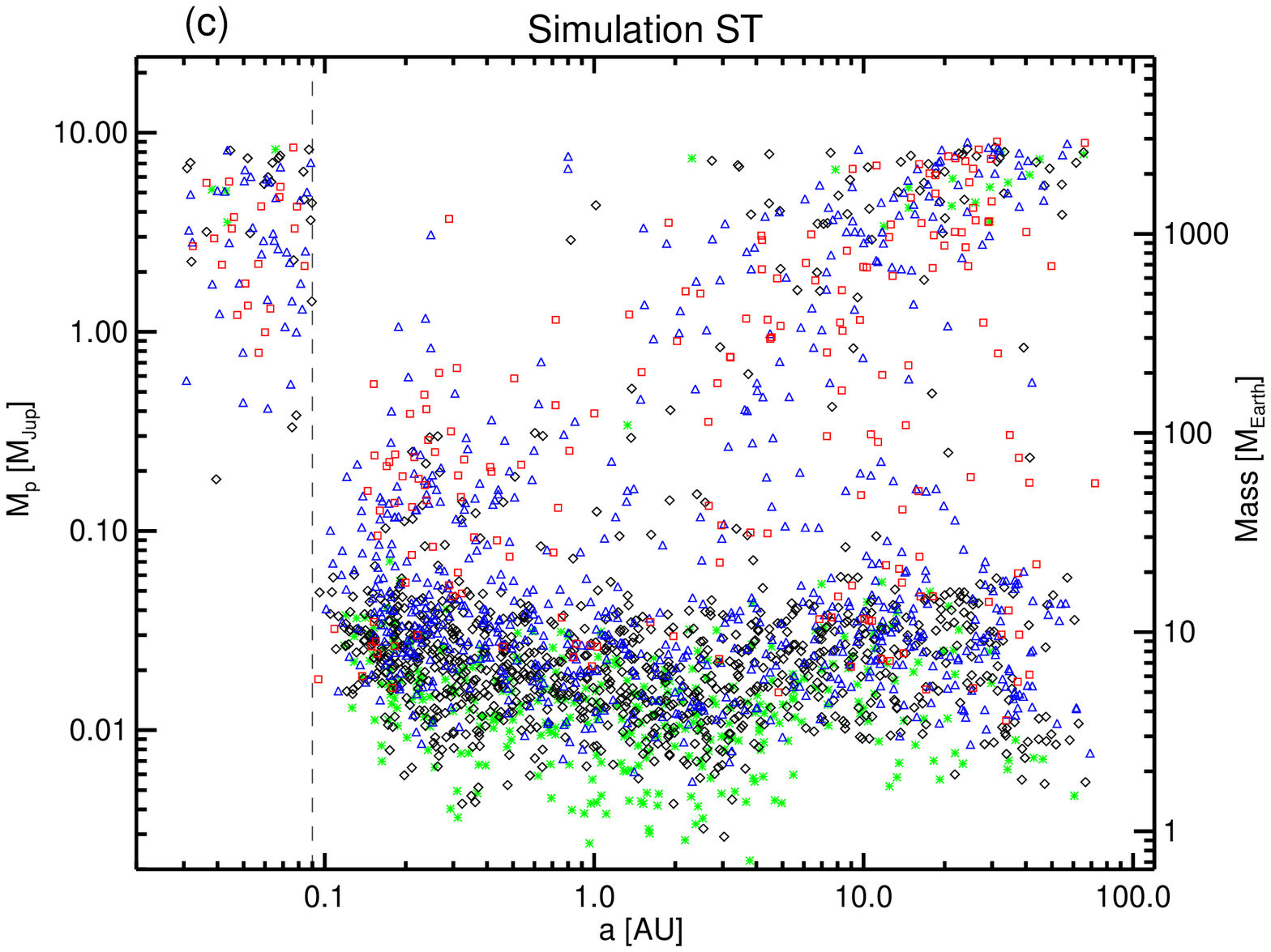}
\caption{Final planet mass versus planet-host separation for simulations: NC (no core formation), DC (``dim cores''), and full core
  growth physics, ST (``standard''). Host star
  metallicities are colour coded).   Note that cores destroy {\em most}
  of far-out giants that would otherwise survive.}
\label{fig1}
\end{figure}

Fig. \ref{fig1} contrasts the final planet
mass versus separation distribution for three population synthesis
calculations of 3000 planet-forming experiments each. Panel (a) shows
simulation NC (``no cores'') in which core formation is artificially suppressed;
panel (b) shows simulation DC (``dim cores'') in which core luminosity is
reduced by an arbitrary factor $10^5$ compared to simulation ST (``standard'')
shown in panel (c) which includes the full planet and core formation physics
of TD model. The main difference of simulation ST from those presented in \cite{NayakshinFletcher15} is in the two parameters that influence grain sedimentation. Whereas the turbulence parameter $\alpha_{\rm d}$ was varied between $10^{-4}$ and $10^{-2}$, here it is set at $10^{-4}$ (cf. Table \ref{tab:1} in appendix \ref{S:popsyn}). In addition, grain breaking velocity is varied between 15 and $30$~m/sec, rather than $5-15$~m/sec range considered in \cite{NayakshinFletcher15}. These choices allow a more rapid core growth, although note that convective grain mixing remains on and affects grain sedimentation significantly.

The colours of the symbols indicate metallicities of  host
stars, as explained in the legend. Panel (d) shows the resulting planet mass
function (PMF) for the three simulations.

As mentioned above, fragments are disrupted when planet's radius $R_{\rm p}$ exceeds its Hills
radius, $R_{\rm H}$. This forms a tidal "exclusion zone" to the left of the thin red
line in panel (a). The exclusion zone boundary depends on planet mass as $a_{\rm exc} \propto M_{\rm p}^{2/3}$. This can be deduced from a simple analytical argument. Pre-collapse gas giants have central temperatures $T_{\rm c}$ less than 2000 K. Planet's density scales as $\rho_{\rm p} \propto M/R_{\rm p}^3 \propto T_{\rm c}^3 M_{\rm p}^{-2}$, where I used $T_{\rm c} \propto M_{\rm p}/R_{\rm p}$, the virial temperature dependence. The planet is tidally disrupted where $\rho_{\rm p} \sim M_*/(2\pi a^3)$, hence the scaling. Migration of post-collapse fragments depends on fragment's mass, of course, and this dilutes the sharpness of the exclusion boundary somewhat.

In simulation NC, nothing is left of the disrupted
fragments when their gas is re-integrated into the disc and then consumed by
the star (or photo-evaporated). The few planets found inside the exclusion zone collapsed before
they entered it. Their protoplanetary disc dissipated away before the planets were pushed all the
way through the inner disc boundary at $0.1$~AU. Most of such close-in
fragments are consumed by the star, although a small number may survive as hot
Jupiters and are shown left of the vertical dashed line. To enhance visibility of those, they are randomly positioned in $\log a$ between $a= 0.03$~AU and $a=0.09$~AU (protoplanetary discs probably have inner boundaries at varying distances from the star due to different strength of their stars' magnetic fields, so this procedure is not entirely unreasonable). For planets outside
the exclusion zone, protoplanetary discs ceased to exist before they migrated
inside the zone. 

Simulation NC shows, in accord with previous results \citep[e.g.,][]{NayakshinFletcher15}, that gas giant planets populating the region inside of the few AU of the host star are mainly metal-rich. This selection is driven by pebble accretion being more vigorous at higher metallicities \citep{Nayakshin15b}. 

In simulation DC, core formation is allowed, and hence here fragment disruption does leave a core behind. It is interesting to note that core masses usually exceed $10 \mearth$ (see panel b of fig. \ref{fig1} and the red histogram in fig \ref{fig:cmf} for more detail). Core's presence does not appear to
affect gas giants that stall outside the exclusion zone (compare
panels a and b) because the cores are very dim in the run DC by construction. Fragments that enter the exclusion
zone before they collapse are disrupted, as in simulation NC. The legacy of disruption in run DC is not just the massive cores. Most of the 
massive cores hold on to the densest part of the fragment in this simulation because the core luminosity is low and hence the gas layers adjacent to the core are quite dense \citep{NayakshinEtal14a,NayakshinFletcher15}. These $\sim$
Neptune to Saturn mass planets continue to migrate through the disc, and are
seen as the nearly vertical "tongue" feature touching the vertical line in
panel (b).

Simulation ST (panel c), performed with our default model for core's luminosity, shows that feedback unleashed by the cores on their host fragments leads to a qualitative change in the resulting planet population. The most frequent type of planet is now a core of a mass $M_{\rm core}\lesssim 10 \mearth$ and these cores are found at all separations explored in this paper. Gas giants are now much rarer, again for all separations. We shall now consider results  from these simulations, and especially ST, in greater detail.

\section{Analysis of simulation ST results}\label{sec:results}

\subsection{Planet mass function}\label{sec:pmf}

\begin{figure} 
\includegraphics[width=\columnwidth]{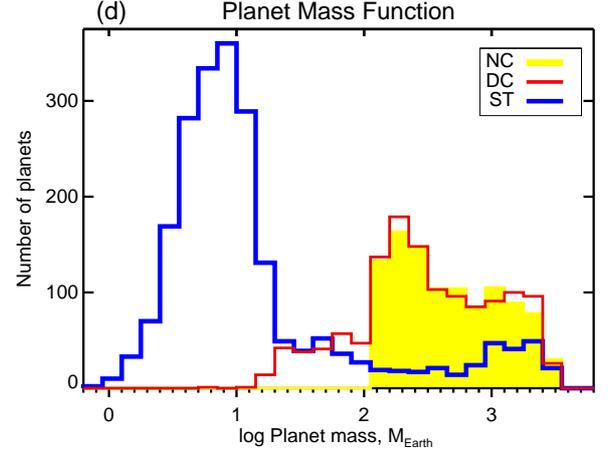}
\caption{The planet mass function (PMF) for the three population synthesis models shown in fig. \ref{fig1}}
\label{fig:pmf}
\end{figure}

Figure \ref{fig:pmf} shows the planet mass function (PMF) for the three simulations. In simulation NC, there are only gas giant planets (of course), with the minimum and maximum masses simply given by the range in the fragment mass in our initial conditions. In simulation DC, fragment disruption extends the PMF towards smaller planet masses, but this extension does not go to masses lower than $\sim 10\mearth$ because the cores are massive and are always surrounded by massive remnant atmospheres. The PMF of simulation ST that includes massive core feedback at the appropriate level is drastically different. The cores of mass $\sim 10 \mearth$ are now the most abundant planets, with very few gas giants surviving. Although disruption of gas fragments always occurs by the means of tides from the stars, comparison of runs DC and ST clearly demonstrates that core's energy output enhances the frequency of fragment disruption significantly. The dominance of massive cores in the PMF of the TD theory is therefore in big part due to feedback released by the cores.

\subsection{Core mass function}\label{sec:cmf}

\begin{figure} 
\includegraphics[width=\columnwidth]{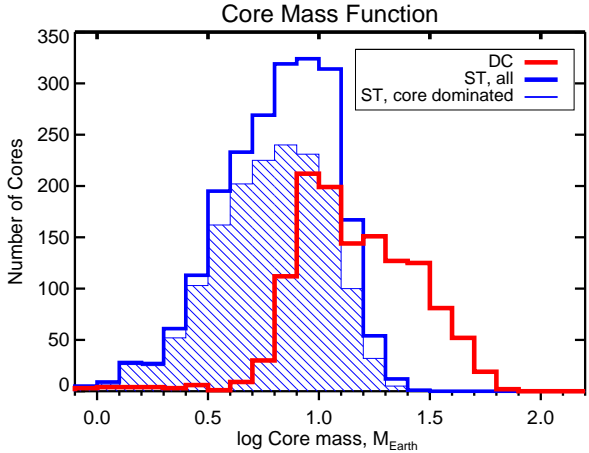}
\caption{The core mass function (CMF) for the population synthesis models DC and ST. The thick solid curves show all cores, whereas the blue shaded histogram shows those that are naked or nearly so. Such cores are found only in simulation ST}
\label{fig:cmf}
\end{figure}

Fig. \ref{fig:cmf} shows just the core mass function (CMF) for the simulations DC and ST (simulation NC has no cores at all by design, so it is not shown in the figure). Further, we divided the core population into those surrounded by a massive gas atmosphere, defined as being more massive than $0.2 M_{\rm core}$, and those with less massive atmospheres ("naked" cores). The hatched histogram in fig. \ref{fig:cmf} shows the population of the naked cores in simulation ST, while the thick solid blue line shows the histogram for all cores in the same run, with or without massive atmospheres. It is clear from these histograms that in simulation ST most of the cores have small atmospheres. The smallest mass cores ($M_{\rm core}\lesssim$ a few $\mearth$) have virtually no atmosphere, but for higher masses the fraction of cores with atmospheres increases. This is in agreement with previous results \citep{NayakshinFletcher15}. For simulation DC, it turns out that all of the cores retain massive atmospheres precisely because they are so dim \citep[see][]{NayakshinEtal14a}, so there is no corresponding naked core population.

The CMF of simulation ST shows a a precipitous drop in the number
of cores that exceed $\sim 10-20\mearth$, in accord with the mass limit given by equation
\ref{Mcrit}. Comparing the CMF of the run ST with that from the run DC we see that there is no corresponding drop in the CMF when core's luminosity is artificially suppressed. There is instead a long tail towards high masses, with the maximum mass of the core assembled in simulation  DC reaching $M_{\rm core}\approx 71\mearth$. Furthermore, as noted above, none of these cores are bare.

We conclude that feedback unleashed by massive cores not only reduces the number of gas giants surviving the disc migration phase
but also regulates the maximum mass of the solid cores, potentially explaining why there are so few cores more massive than $\sim 20 \mearth$ \citep{MayorEtal11}. 
In \S \ref{sec:special} I argue that this explanation is better consistent with the data than that given by CA theory.

\subsection{Cold disruptions: "deficit" of observed gas giants}\label{sec:far_out}

A popular argument against GI model for planet formation, and by extension against TD as well, is that gas giant planets found at separations greater than 10 AU are exceedingly rare in the direct imaging surveys. \cite{BillerEtal13}, for example, finds that no more than a few \% of stars host $1-20\mj$ companions with separations in the range $10 - 150$~AU. Rapid inward migration may be expected to remove most of such objects even if they did form at tens of AU. However,
the fraction of those that survive in our population synthesis simulations {\em without core feedback} is  still too high. In simulation DC, for example, 15.6\% of the initial gas fragments survive in orbits beyond 10~AU. When the feedback is taken into account, more than two thirds of gas giants that are stranded outside the exclusion zone in simulations NC and DC are disrupted by the high luminosity cores forming inside.  In simulation ST, only 4.7\% of initial fragments remain at orbits wider than 10 AU by the end of the simulations. This is probably consistent with the observational limits, especially if (a) protoplanetary discs smaller than $\sim 50$~AU never fragmented -- these stars would be planet free in the TD model, and would dilute the planet-bearing systems we discuss here; (b) companions at larger separations or close passages of other stars in star clusters eject some of the left-over giants from the systems. 

\cite{RiceEtal15} argued that the frequency of fragment formation via disc fragmentation  can be constrained by  an additional argument. They find that about $\sim 5$\% of their fragments, initially located at $a > 50$~AU, get scattered on very close orbits by distant companion stars. The authors then show that these fragments would have a very different metallicity distribution than the planets actually observed at small separations. This is clearly correct, however,  \cite{RiceEtal15} simulations begin when the gas disc dissipates, that is roughly at the end of the simulations presented here, and concern only the clumps that are stranded on distant orbits. Most of our fragments ($\sim 95$\% in simulation ST) have migrated due to the gas disc torques into the inner disc or have been disrupted by the feedback by that time. Hence, the frequency of occurrence of objects scattered by the secondary's N-body torques to small separations \citep[as calculated by][]{RiceEtal15} in simulation ST would be at most about $0.05 \times 0.05 = 2.5 \times 10^{-3}$, which is too low compared with the population of planets that migrate there directly during the gas disc migration phase. Hence, \cite{RiceEtal15} arguments do not contradict our scenario at all.

Furthermore, disruption due to cores is especially severe for moderately massive fragments. For fragments with mass $M_{\rm p} \le 3\mj$, for example, only 1.5\% of the initial fragments survive at large separations. If only moderately massive fragments are hatched by the disc at beyond tens of AU, then the simulation ST is consistent with the observational limits on the frequency of directly imaged gas companions {\em even if} several fragments are born per each disc\footnote{This is especially so since we simulate the last clump in the dispersal phase of the disc, when the disc starts to run out of mass. The clumps found beyond 10 AU in our simulations are those for which the disc was removed particularly rapidly so that they got stranded there. The earlier generations of clumps, formed in the phase of the disc when it was still accreting gas from the envelope, 
would never be in that regime of disc running out of mass. These early fragments are very likely to migrate in rapidly and contribute to the observed populations of close-in super-Earth planets rather than gas giants beyond 10 AU.}. 

TD with feedback hence predicts a significant population of cold massive solid cores, and downsized giants ($\sim$ sub-Saturn planets) more generally, formed rapidly. Existence of such cores is an observational test of this theory, as opposed to the currently dominant view that fragmentation of self-gravitating discs is rare.

\begin{figure}
\includegraphics[width=0.95\columnwidth]{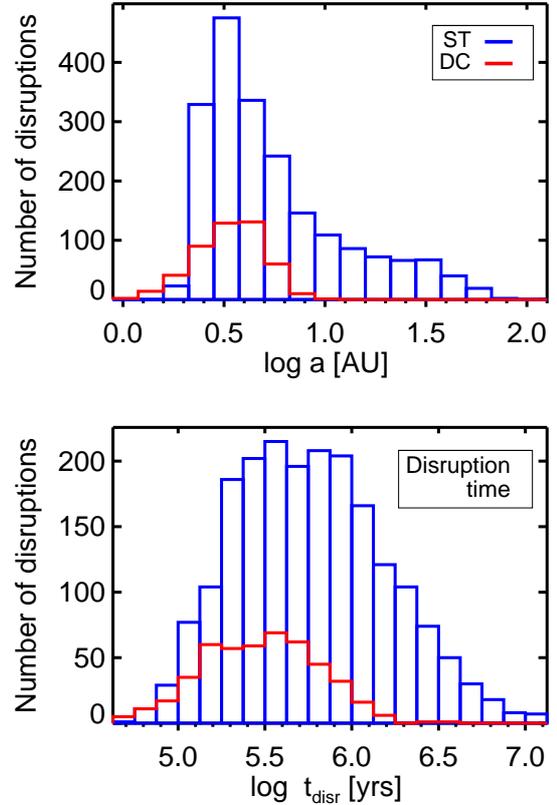}
 \caption{Top panel: host-fragment separation at the time of fragment disruption
   for simulations DC and ST. Bottom panel: same but for the time of fragment
   disruptions.}
   \label{fig3}
 \end{figure}

\subsection{Massive and distant cores, unexpectedly young}\label{sec:HL}

The population of the distant young cores that must be produced when would-be-giant planets die at large separations is discussed in this section.
Fig. \ref{fig3} shows the locations (top panel) and the times (bottom panel) of fragment
disruptions in simulations ST and DC. We already saw that luminous cores force many more
disruptions than do ``dim cores''. This is why the blue histograms contain many more disruption events than do the red ones. 
The top panel of fig. \ref{fig3} shows that in addition to that,
disruptions at large ($a > 10$~AU) separations occur only in simulation ST. This demonstrates that feedback from massive cores not only increases the number of disruptions but it also changes the conditions under which these disruptions occur. 

Consulting panel (c) of fig. \ref{fig1}, we notice that the cold disruptions produce massive cores with mass between that of a Super Earth and Neptune, although some cores manage to hold on to more massive gas atmospheres despite the host fragment's disruption and reach $\sim$ Saturn mass. The bottom panel of fig. \ref{fig3} shows that many of the cold disruptions occur at a young age of the system, at a fraction of a Myr. This is interestingly different from the CA theory where massive cores assembled at large distances take $10-100$ Myr to grow \citep{KobayashiEtal11,KB15}.

In sections \ref{sec:HLTAU} and \ref{sec:UN},  I shall argue that such cold disruptions induced by massive core feedback may offer natural explanations for assembly of Uranus, Neptune, and the suspected planets in the HL Tau protoplanetary disc.

Furthermore, here we explored the situations where the starting embryo positions are not larger than $a=105$, a plausible but not a necessary choice. 
Simulations of protoplanetary discs that include its formation from larger scales show that these discs may well hatch fragments at larger radii \citep[e.g.,][]{VB06,Vorobyov13c}. It therefore seems plausible that massive
cores formed in TD by self-destruction of their fragments could be formed in
young protoplanetary discs beyond 100 AU. In appendix \ref{S:cold} and in Fig. \ref{Sfig_a150}, I show an
example run in which a fragment born at $a=150$~AU self-disrupts and leaves a
core of mass $M_{\rm core} =5.1\mearth$ stranded at $a= 112 AU$.

\subsection{Large debris discs}\label{sec:debris}

Not all large solids are expected to condense into the core if the fragment is rotating, which is always the case in the simulations of fragmenting gravitationally unstable protoplanetary discs  \citep[e.g.,][]{BoleyEtal10,Nayakshin11a}. Therefore, it is reasonable to assume that large $\sim 100$~km sized bodies may form inside fragments 
and be orbiting within it due to excess angular momentum prior to disruption. \cite{NayakshinCha12} showed that in this case disruptions of pre-collapse fragments leave behind not only the solid cores, but also rings of smaller solids. These rings are morphologically similar to the Asteroid and the Kuiper belts in the Solar System and the debris discs elsewhere \citep[e.g.,][]{Wyatt08}. Such post-disruption rings of solid bodies is the TD alternative to the planetesimal discs postulated by the Core Accretion \citep{Safronov72}.

I defer detailed discussion of the debris disc properties formed in this model to a future publication (Fletcher \& Nayakshin, in preparation). Nevertheless, it is relevant to note here the role of feedback in formation of these rings. Because feedback produced by the solid cores helps to disrupt the fragments at all radii $a > 1$ AU (top panel of fig. \ref{fig3}), we find debris discs of all sizes, compatible with the broad range of size in observed debris discs. Core feedback is hence instrumental to understanding the debris discs in the context of TD as well.

\subsection{Metal Monsters}\label{sec:monsters}

\begin{figure} 
\includegraphics[width=1.05\columnwidth]{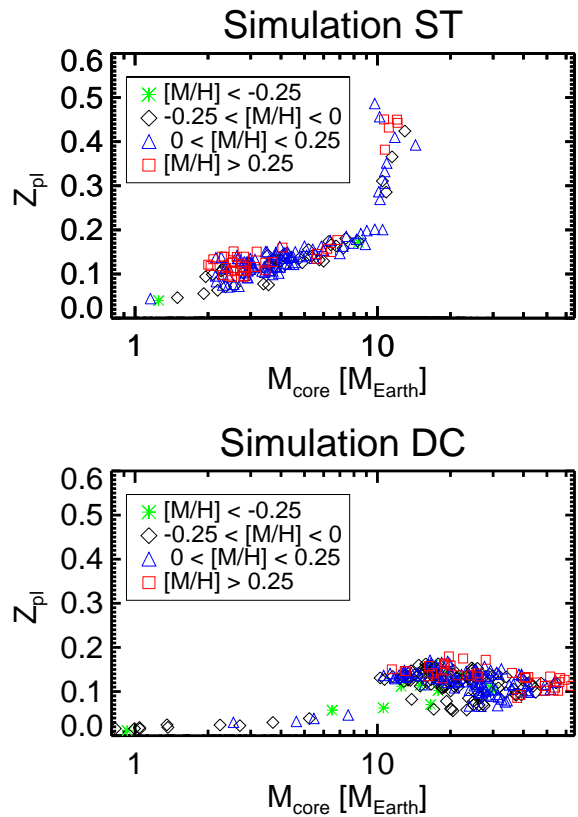}
\caption{Internal metallicity of gas giant planets for $M_{\rm p} > 1 \mj$ as the function of their core mass for simulations ST and DC. 
Only a subset of the planets, selected randomly, is shown for clarity.}
\label{fig:zmon1}
\end{figure}

Fig. \ref{fig:zmon1} plots the planet's metallicity, $Z_{\rm pl} = M_{\rm Z}/M_{\rm pl}$, where $M_{\rm Z}$ is the total weight of metals (elements heavier than Helium) inside a planet of mass $M_{\rm p}$, versus the planet's core mass for planets that were not disrupted\footnote{These are the gas giants planets, of course. The planets less massive that $\sim$ Saturn in TD belong to the disrupted population. These have even higher metallicities due to the presence of a massive core and their remaining gas envelopes being more metal rich than the original envelopes. See fig. 17 in \cite{NayakshinFletcher15}. To avoid these disrupted planets we include only $M_{\rm p} > 1 \mj$ giants in the figure.}. The top panel shows simulation ST and the bottom one shows simulation DC. The colours of the symbols mark the four different metallicity bins, as detailed in the legend.

There is a striking difference in $Z_{\rm pl}$ vs $M_{\rm core}$ trends for the planets assembled with (simulation ST) and without (simulation DC) core feedback. The metallicities of the planets assembled without feedback first rise with $M_{\rm core}$ and then drop a little (the bottom panel). The rising trend at low $Z_{\rm pl}$ is due to metal-limited growth of cores for such low $Z_{\rm pl}$: more metals implies faster grain sedimentation and a higher $M_{\rm core}$. For most of the cores in simulation DC, planet's metallicity is spread between 0.05 and $\sim 0.15$.

Inclusion of core feedback appears to drastically change this picture. First of all, feedback self-limits mass of the cores, as explained in \S \ref{sec:cmf}. This is why $M_{\rm core}$ values are all below $\sim 15\mearth$ in the top panel. Furthermore, the $Z_{\rm pl}$ vs $M_{\rm core}$ relation is steeper and nearly monotonic: the higher the core mass, the higher the planet's metal content. There is also a nearly vertical sequence of points towards surprisingly high metallicities up to $Z_{\rm pl}\approx 0.5$, taking off at core mass $M_{\rm core}\sim 10 \mearth$.

\begin{figure} 
\includegraphics[width=1.05\columnwidth]{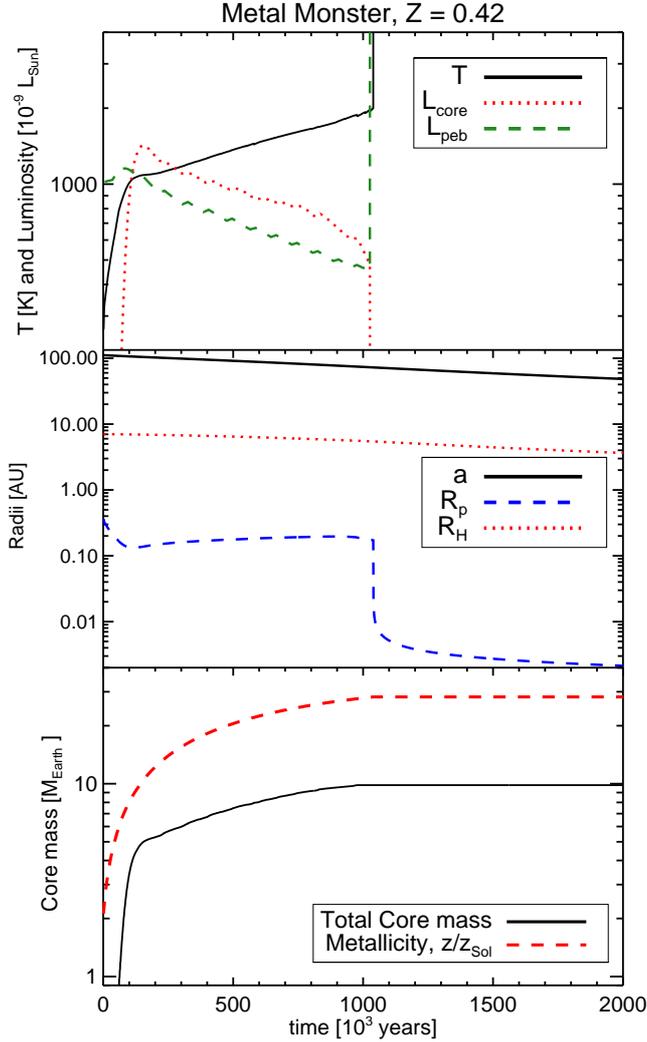}
\caption{Similar to fig. \ref{fig2}. Evolution of a gas fragment versus time in a simulation that forms a metal monster planet. Note that unlike fig. \ref{fig2}, the core's luminosity drops in this simulation shortly after the fragment starts to expand. This allows the fragment to hover near marginal stability while it is fed pebbles from the outside, and become very metal rich.}
\label{fig:zmon2}
\end{figure}

The results for simulations ST are best understood as a time sequence of $M_{\rm core}$ and $Z_{\rm pl}$. Initially, as the planet accretes pebbles and the core grows, these two quantities co-evolve to larger values on parallel tracks. However, when $M_{\rm core}$ reaches the mass limit $M_{\rm sd} \sim 5-10 \mearth$ given by equation \ref{Mcrit}, its energy output starts to affect the core strongly. I find that one of the following two outcomes happen after this point.

For most of the fragments, the core luminosity continues to increase (as in panel b of fig. \ref{fig2}), and this puffs up the fragment. Eventually it is disrupted. This is why the cores in the top panel do not make it to higher masses, unlike the feedback-free case (bottom panel). 

However, there is a small number of cases when there is a certain degree of self-limitation in the core growth, so that the fragment is not immediately disrupted. As the fragment expands, core growth rate drops because grain velocity is the sum of the (negative) sedimentation velocity and the gas (positive in this case since the fragment is expanding) velocity. In this situation the opposite terms in equation \ref{etot1}, $L_{\rm core}$ and $L_{\rm peb}$, nearly balance each other and the fragment's radius evolution stagnates despite pebble accretion. The fragment then becomes much more metal rich than expected based on the analytical model of pebble accretion by \cite{Nayakshin15a}, in which cores were not taken into account. In that theory the fragments should collapse once $Z_{\rm pl}\sim 0.1$ is reached, and it fits well the numerical results for the majority of planets in both panels of fig. \ref{fig:zmon1}.

The exceptional systems that reach much higher $Z_{\rm pl}$ on the vertical track exist due to core feedback. Their evolution is exemplified by fig. \ref{fig:zmon2} which shows the evolution of the fragment which reached very high metallicity by the time it collapsed, $Z_{\rm pl} = 0.42$. Focusing on the red (dotted) and the green (dashed) curves in the top panel of fig. \ref{fig:zmon2}, showing $L_{\rm core}$ and $L_{\rm peb}$, respectively, we see that $L_{\rm core}$ indeed drops soon after the two curves intersect. This is connected to the fact that the core accretion rate dropped significantly after the feedback started to inflate the planet (as can be seen from the abrupt change in the slope of $M_{\rm core}$ vs time curve in the bottom panel of the figure). 

I find that the fraction of gas giant planets that follow the nearly vertical evolution line in the $Z_{\rm pl}$ vs $M_{\rm curve}$ plane depends on several somewhat poorly known parameters of my numerical model, such as $t_{\rm kh}$ that determine the timescale on which cores release their heat of accretion, disc pebble fraction $f_{\rm p}$, grain breaking velocity, $v_{\rm br}$ and the turbulence in the fragment, $\alpha_{\rm d}$ (see appendix \S \ref{S:popsyn}). Therefore, the fraction of such "metal monster" gas giants cannot be yet reliably calculated, except to say that such planets should be rare, $\simlt 10$\%. 

I propose that the surprisingly high metallicity planet CoRoT-20b \citep{DeleuilEtal12} formed as described in this section.

\section{Discussion}\label{sec:discussion}

\subsection{Main result and comparison to previous work}

Here, no major changes in the TD model for planet formation were introduced compared to a recent \cite{NayakshinFletcher15} study, except for relaxing somewhat factors controlling core growth by grain sedimentation (reducing turbulence and increasing the grain breaking velocity, see \S \ref{S:cores} and table \ref{tab:1}). This led to a somewhat more rapid massive core assembly, and put the effects of feedback by massive cores on their host fragments into a sharper focus. Cores more massive than $\sim 5 \mearth$ (eq. \ref{Mcrit}) affect their host gas fragments significantly by releasing enough energy to puff the fragments up. This leads to more frequent tidal disruptions of these fragments, especially at large separations from the star, and explains naturally why the TD mass and core function is strongly cut off above the mass of $\sim 20\mearth$.

Now I discuss the most interesting observational implications of these results.

\subsection{The special role of $M_{\rm core}\sim 10 \mearth$}\label{sec:special}

Small, core-dominated planets, $R_{\rm p} \lesssim 4 R_\oplus$, are very ubiqutous \citep{HowardEtal12}. In terms of mass, this size range translates  to somewhere around $M_{\rm p}\sim 20\mearth$, above which the planet mass function (PMF) drops sharply \citep{MayorEtal11}. This indicates that cores with mass $M_{\rm core}\sim 10 \mearth$ have a special role to play in planet formation. 

CA model suggest that this special role is in building massive gas giants: cores more massive than $M_{\rm crit}$ accrete gas rapidly \citep[e.g.,][]{PollackEtal96}. However, the critical core mass can actually vary substantially depending on the planetesimal or pebble accretion rate \citep{Stevenson82,IkomaEtal00,LJ14}, location in the disc \citep{Rafikov06},  dust opacity \citep{IkomaEtal00}, and the metallicity of the atmosphere \citep{HoriIkoma11}. The range here is from a few $\mearth$ to $\sim 50\mearth$. It is not obvious that such a broad range of critical core masses can yield as sharp a cutoff in the PMF as observed. Furthermore, if cores that grow more massive than $10-20\mearth$ rapidly become giant planets by gas accretion, when where are\footnote{The author thanks Vardan Adibekyan for raising this question.} these giants? This scenario would predict roughly $\sim 10$ times more gas giants than observed (cf. the relative numbers of planets discussed in the next section), given the ubiquity of massive cores.

The explanation advanced in this paper for the special place of $M_{\rm core}\sim 10 \mearth$  cores in planet formation paints the cores not as nurseries of gas giant planets but as sources of powerful feedback that leads to self-destruction of most gas fragments. This negative role of cores is consistent with that played by densest parts of all larger astrophysical systems. Stars, giant molecular clouds, star clusters, and galaxies all loose mass because of energy released in the densest parts of these systems.

The critical core mass derived here (eq. \ref{Mcrit})  does not depend strongly on the host star or fragment metallicities, or the distance to the host star, and does not seem to contradict any constraints from the observations of exoplanets at close separations. It appears both more robust and more logical than the one offered by the CA scenario.

\subsection{The rarity of gas giant planets}\label{sec:rarity}

Gas giant planets are very rare compared to core-dominated planets (those from Earth mass to $\sim$ Neptune mass). In a recent very careful analysis of {\em Kepler}-detected giant planet candidates, Santerne et al (2015, submitted) have shown that more than 50\% are false positives, such as eclipsing binaries. This corrects downward the giant count, so that the overall frequency of gas giant planets with orbital period less than 400 days is estimated to be only $4.6\pm 0.6$\%. This is over 10 times less frequent than planets with radii between 1 and $4 R_\oplus$ \citep{SilburtEtal15}. In TD, this result is not surprising giving that most migrating gas fragments are disrupted and leave behind a massive core.

In addition, directly imaged surveys conclude that $M_{\rm p} > 1 \mj$ planets are hosted by no more than a few \% of nearby stars \citep{BillerEtal13,BowlerEtal15}. This is thought to indicate that gas disc fragmentation at $\simgt $ tens of AU is exceedingly rare \citep{RiceEtal15}. However, these observations do not directly tell us the number of gas fragments born by GI in the outer disc; they only constrain the fraction of giants that survive to the present day. Migration of planets to closer separations was found to take as little as $\sim 10^4$ years in numerical simulations by numerous authors \citep[e.g.,][]{VB06,BoleyEtal10,BaruteauEtal11,ChaNayakshin11a,MichaelEtal11,MachidaEtal11}.

Further, here it was shown that disruption of fragments by feedback from massive cores also destroys many of the would-be giant planets. In \S \ref{sec:far_out} I showed that TD with feedback is at least qualitatively consistent with the low observed frequency of gas giant planets in directly imaged surveys \citep{BillerEtal13,BowlerEtal15}. 

 In detail, however, I find that theoretical predictions on number of giants stranded at large separations depend sensitively on the mass function of gas fragments born in the outer disc. Fragments more massive than $\sim 4-5\mj$ are unlikely to make massive cores and hence they are not likely to be tidally disrupted. If fragment mass function is "bottom-heavy", with most fragments at $\sim 1\mj$, then observational constrains appear to be easily satisfied; however, if most fragments were more massive than a few $\mj$ then the model would probably over-produce "cold" giants (this can be qualitatively deduced from fig. \ref{fig1}c). More work, e.g., 3D simulations to understand fragment formation and their typical properties better, is needed for a quantitative comparison of TD theory with data at large separations.


\subsection{The "impossible" planets in HL Tau}\label{sec:HLTAU}


Gas fragment disruption by massive cores leaves the perpetrators behind. These cores can sometimes be observable. I argue that one such case may be the dust-depleted rings in HL Tau, and the other are the outer Solar System (\S \ref{sec:UN}).  

HL Tau is a young ($\sim 1-2$ Myr old) star located $\sim 140$~pc away from the Sun and is the brightest protoplanetary
disc in millimetre radio emission \citep{AndrewsWilliams05}. Atacama Large Millimetre/Submillimetre Array (ALMA) has recently observed HL Tau and produced the first ever {\em picture} of planet formation disc. The image of HL Tau shows a number of circular depressions in the dust emissivity of the
disc. \cite{DipierroEtal15} reproduces the observed ALMA image in tantalising detail by having three planets with mass between $\sim 0.2$ to $\sim$ 0.5 $\mj$, orbiting the star at $a=13$, 32 and 69 AU, respectively. The planets affect the surrounding gas disc weakly, but force mm-sized grains (to which ALMA is sensitive) away from their orbits, producing the observed dark rings.

The actual masses of planets in HL Tau may be lower that the $\sim$ Saturn masses inferred by  \cite{DipierroEtal15}. As shown by \cite{BaruteauEtal11}, 
Saturn or higher mass planets migrate through an outer self-gravitating disc in $\sim 0.03$~Myr or less. This is consistent with many other studies \citep{BoleyEtal10,ChaNayakshin11a,MichaelEtal11}. The planet migration time scales as $t_{\rm migr} \propto M_{\rm p}^{-1}$ in the type I regime. Hence the planets of Neptune mass would migrate $\sim$ an order of magnitude slower, and would be more likely to have been detected by ALMA.

Additionally, at least in the Core Accretion theory, planets as massive as $\sim 0.2\mj$ should be accreting gas very rapidly in the HL Tau rings because the observed gaps are in the dust grains not in the gas \citep{DipierroEtal15}; gas surrounds the planets. Mass doubling time for the parameters appropriate for HL Tau disc can be estimated at $\sim$ a few $\times 10^4$ years. These planets should run away in mass into $\simgt$ Jupiter mass planets. Gas accretion onto the planet should make the planets bright enough to have been detected by ALMA as brighter spots in the rings.

HL Tau planets properties are also inconsistent with the simplest, that is migration-free, variant of the Gravitational Instability model \citep[e.g.,][]{Boss97}. The planets are too close-in (especially the one at $a\approx13$~AU), and are too low mass. The minimum (Jeans)
mass of fragments made by GI is about half of Jupiter mass \citep{BoleyEtal10} if not more \citep{ForganRice13}. 

In contrast, here it was shown that TD naturally produces $\sim$ Neptune mass planets in the relevant separation, time and mass ranges (see \S \ref{sec:HL}). Such planets do not accrete gas from the disc, at least not until their cores cool down, which is probably longer than the disc lifetime in most systems. 

A planet mass nearer to Neptune mass would yield much longer migration time scales, closer to the expected age of HL Tau. The smaller planet mass would  also be consistent with the $\sim$ Neptune mass limits earlier obtained for the embedded HL Tau planets by \cite{TamayoEtal15}. This would not necessarily contradict \cite{DipierroEtal15} results.  Due to numerical constraints, their simulations were run for only $\sim 5,000$ years, which is much shorter than the likely HL Tau age. Neptune or somewhat smaller mass planets may well open sufficiently deep depressions in the dust distribution in simulations ran for longer and/or with lower artificial viscosity.

In appendix \ref{S:cold}, an even more distant massive core formation was demonstrated. TD hence can produce massive young cores orbiting the host star well beyond 100 AU. TD hence predicts that dust rings in discs even larger than those detected by ALMA in HL Tau may be common place in young protoplanetary discs.

\subsection{Uranus, Neptune and the outer Solar System}\label{sec:UN}

Uranus and Neptune, located at $a\approx 19$ and 30 AU, respectively, have masses $M_{\rm p} = 14.5$ and $17 \mearth$. Even for Uranus with its less distant location, \cite{PollackEtal96} found core growth time of 16 Myr, which is too long to explain the presence of Hydrogen on that planet. More recent models find much higher solid accretion rates \citep[e.g.,][]{LambrechtsJ12}, cutting the growth time for Uranus and Neptune's cores significantly. However, \cite{HB14} show that for these higher accretion rates the issue is in understanding why these planets accreted just a little gas, e.g., why gas accretion stopped just before the runway gas accretion phase commenced, and that formation of such planets in CA is still not trivial.

In the model presented here, massive cores of $\sim 10-15\mearth$ with Hydrogen atmospheres of a few $\mearth$ form quite frequently (see fig. \ref{fig:cmf}). The special meaning of this mass range is explained rather naturally by eq. \ref{Mcrit}. The cores do not accrete more gas from the disc easily since their luminosity is significant.

As suggested earlier by \cite{HW75,NayakshinCha12}, the outer Solar System, including the Kuiper belt, may be rather naturally formed due to self-induced disruptions of two or more gas fragments of a Jupiter mass.  More generally, here it was shown that self-disruption of planets at large distances makes debris discs with radii of tens and even hundreds of AU in the context of TD, as required to explain the observed large debris rings \citep[e.g.,][]{Wyatt08}.

\subsection{Metal monster planets}\label{sec:zmon}

Planets in general, and gas giant planets in particular, are over-abundant in metals compared to their host stars \citep{SG04,MillerFortney11}. In TD scenario, such over-abundance of metals arises from accretion of pebbles onto pre-collapse gas fragments \citep[cf. \S 8 and fig. 17 in][]{NayakshinFletcher15}. \cite{Nayakshin15a,Nayakshin15b} argued that gas fragments accrete $\sim 10$\% of their mass in pebbles before they collapse by H$_2$ dissociation.

Strong feedback by the core puffs the fragments up, hence requiring more pebble accretion for collapse. In \S \ref{sec:monsters} it was shown that the more massive the core is, the more metal rich the gas giant planet is, in general. In addition, a branch of even more highly metal-enriched planets was found, with metallicities $Z_{\rm pl}$ up to extreme values of $\sim 0.5$ (see the top panel of fig. \ref{fig:zmon1}). These  "metal monsters" are rare in the current simulations (a few \% of the survived gas giant population), and more detailed fragment and disc models are needed to nail down their expected frequency. 

Nevertheless, it is interesting to note that there may already exist observational counterparts to the metal monsters. CoRoT-20b is a $M_{\rm p}\approx 4.2 \mj$ gas giant that
has been estimated to contain $\sim 800 \mearth$ (which is $\sim
60$\% by weight!) of metals if they are all locked in the core, or perhaps
$\sim 300$ to $400\mearth$ if the metals are in the envelope
\citep{DeleuilEtal12}.  Its metal abundance is hence $0.25 \le Z_{\rm pl} \le 0.6$. 
Furthermore, many of the strongly irradiated gas giant planets are inflated \citep{MillerFortney11}, but others are not. It is possible that those that are not inflated despite being irradiated by high fluxes contain more metals than assumed in the models. This implies that some of such non-inflated but very hot Jupiters may have very high metallicities.

\subsection{Uncertainties and future work}

The sources of uncertainties in this work come from several factors. First of all, it is desirable to model the internal evolution of the fragment better by using a fully fledged planet/stellar evolution code, as done, by, e.g., \cite{VazanHelled12}. This may affect the maximum mass of the cores made inside the fragments. It is also desirable to model the "solid" core explicitly via the same stellar evolution code rather than via a constant density approximation to be able to calculate the energy release by the core better. This is a very difficult task at the moment since opacity and equation of state for hot super-Earth mass cores are not well understood yet \citep{StamenovicEtal12}.

Grain micro-physics is another worry. Three grain species are included here, but in reality there are many more, and there is a possibility that individual grains are composed of a number of different species.

Fragment-disc interactions are another source of model dependencies. In particular pebble accretion rates, are modelled in the "Hill's regime", e.g., when all pebbles (grains with Stock's number $\sim 1$) entering the Hill's sphere of the planet are accreted \citep{LambrechtsJ12}. This may not be a good approximation if the planet itself is migrating rapidly through the disc. A 3D hydrodynamics study of this issue is required.

These and many other modelling uncertainties are clearly a worry. However, the critical self-disruption mass limit (eq. \ref{Mcrit}) is derived based on a simple energy argument, so it is hard to see how it may be {\em very} sensitive to these effects. Furthermore, another major set of results, the planet correlations with host metallicities \citep[which were explored in great detail in][]{NayakshinFletcher15} are likely to be robust because they arise from "common sense" trends, such as higher pebble accretion rate onto the fragments in high metallicity protoplanetary discs, which are expected to hold in 1D or 3D simulations.

\section{Conclusions}

Here it was suggested that the role of massive cores in gas giant planet formation is not at all positive. By unleashing energetic feedback onto their host fragments in the first $\sim 0.1-1$ Million years of their life, when the fragments are most vulnerable to stellar tides of the host star, massive cores ensure that they are the dominant population of planets. However, the cores are also victims of their own success: their mass cannot exceed $\sim 10-20\mearth$ because their hosts cannot withstand stronger feedback by more massive cores, and get disrupted before the cores could grow more massive. There is a number of observational predictions that distinguish TD from CA scenario and hence I hope that these two theories will be observationally differentiated in a very near future.

\section*{Acknowledgements}

Theoretical astrophysics research at the University of Leicester is supported
by a STFC grant. This paper used the ALICE High Performance Computing Facility at the University of Leicester. The authors is thankful to the anonymous referee whose comments improved the paper significantly.

\bibliographystyle{mnras}



\appendix

\section{Numerical Methods}\label{S:Appendix}

\subsection{Description of a single planet formation experiment}\label{S:exp1}

Each planet formation experiment presented here models the last stages of the
protoplanetary disc evolution, in which the disc around a protostar of mass
$M_* = 1 M_\odot$ stops receiving mass from an external infalling gas
envelope.  During the simulations, the disc mass monotonically decreases due
to accretion of gas onto the star and photo-evaporation of the disc
material. The disc photo-evaporation prescription is the sum of internal and
an external photo-evaporation terms (see below), normalised so that the mean
disc lifetime is $\sim 3$ Million years, as observed \citep{HaischEtal01}.

In parallel to this, the evolution of a gas fragment born in the disc at
separation $a\sim 100$~AU is followed. The fragment exchanges angular momentum
with the disc via gravitational torques and migrates inward. The simulation
stops when either the fragment migrated all the way through the disc to the
disc inner boundary at $R=R_{\rm in}$, or both the disc and the fragment stop
evolving. In the latter case, by the end of the simulation the disc is
dissipated away and the planet has either collapsed (became a gas giant
planet) or was downsized to a smaller planet. Typically, the simulations span
from a fraction of 1 Million year to a little over 10 Million years. A single
run takes between a few hours to a few days on a single CPU. I use the
University of Leicester and UK national super-computers to run tens of
thousands of models (see Acknowledgement).

Although it is likely that a number of fragments are born in the
self-gravitating protoplanetary disc, especially in the earlier more massive
phase, only one fragment per disc is simulated at this time.

\subsection{Disc evolution model}\label{S:disc}

The protoplanetary disc is treated in the 1D azimuthally symmetric
approximation, on a logarithmic radial grid extending from radius $R_{\rm
  in}=0.08$~AU to $R_{\rm out} = 400$~AU. The disc has an initial surface
density profile $\Sigma_0(R)\propto (1/R) (1 - \sqrt{R_{\rm in}/R})
\exp(-R/R_{\rm disc})$, where $R_{\rm disc} = 100$~AU is the disc radial
length scale. The normalisation for $\Sigma_0(R)$ is set by specifying the
total initial disc mass, $M_{\rm d}$, integrated from $R_{\rm in}$ to $R_{\rm
  out}$.  To evolve the disc, the time-dependent standard accretion disc
equations of \cite{Shakura73} are solved. For the surface density equation,
the additional torque of the planet on the disc and the disc photo-evaporation
terms are included:
\begin{equation}
  \frac{\partial\Sigma}{\partial
  t}  = \frac{3}{R}\frac{\partial}{\partial R}
  \left[R^{1/2}\frac{\partial}{\partial R}(R^{1/2}\nu\Sigma)\right]
  -\frac{1}{R}\frac{\partial}{\partial
  R}\left(2\Omega R^2\lambda\Sigma\right) - \dot \Sigma_{\rm ev}
\label{eq:diffplanet}
\end{equation}
where $\Sigma(R)$ is the disc surface density, $\Omega(R) = \sqrt{GM_*/R^3}$
is the Keplerian angular frequency at radius $R$, viscosity $\nu = \alpha_{SS}
c_s H$, where $c_s$ and $H$ are the midplane sound speed and the disc height
scale, $\lambda=\Lambda/(\Omega R)^2$, and $\Lambda$ is the specific tidal
torque from the planet, which switches from type I to type II prescription
following the results of \cite{CridaEtal06}. The expression for both type I
and type II $\lambda$ are standard [see \cite{Nayakshin15c}], but type I
migration regime timescale is further multiplied by a factor $1 \le f_{\rm
  migr}$ as in \cite{IdaLin04b}.

The last term in equation \ref{eq:diffplanet} is the disc photo-evaporation
rate, which is the sum of internal UV and X-ray terms
\citep{NayakshinFletcher15}, e.g., due to the host's star radiation, and an
external irradiation term. The latter is given by expressions from
\cite{Clarke07} multiplied by 0.01. Following \cite{MordasiniEtal12}, we
introduce a Monte Carlo random variable $0.02 < \zeta_{\rm ev} < 3$, which is
the photo-evaporation rate multiplier. The upper and lower limits to
$\zeta_{\rm ev}$ are found such as to fit the mean observed protoplanetary
disc lifetime of $\sim 3$~Myr \citep{HaischEtal01}.

The disc viscosity follows the classical \cite{Shakura73} prescription, but
with the coefficient $\alpha = \alpha_0 + \alpha_{\rm sg}$, being a sum of a
constant, $\alpha_0$, and the self-gravity term that depends on the local disc
Toomre's parameter, $Q$:
\begin{equation}
\alpha_{\rm sg} = 0.2 {Q_0^2 \over Q_0^2 + Q^2}\;.
\label{alpha_sg}
\end{equation}
where $Q$ is the local Toomre's parameter. The parameter $Q_0$, setting the
transition from a viscosity driven by gravito-turbulence \citep{Rice05} to the
non self-gravitating one, is set to $Q_0=1.5$ here. $\alpha_0$ is a
Monte-Carlo variable (cf. Table S1).

\begin{figure}
\centering
\includegraphics[width=3.in,angle=0]{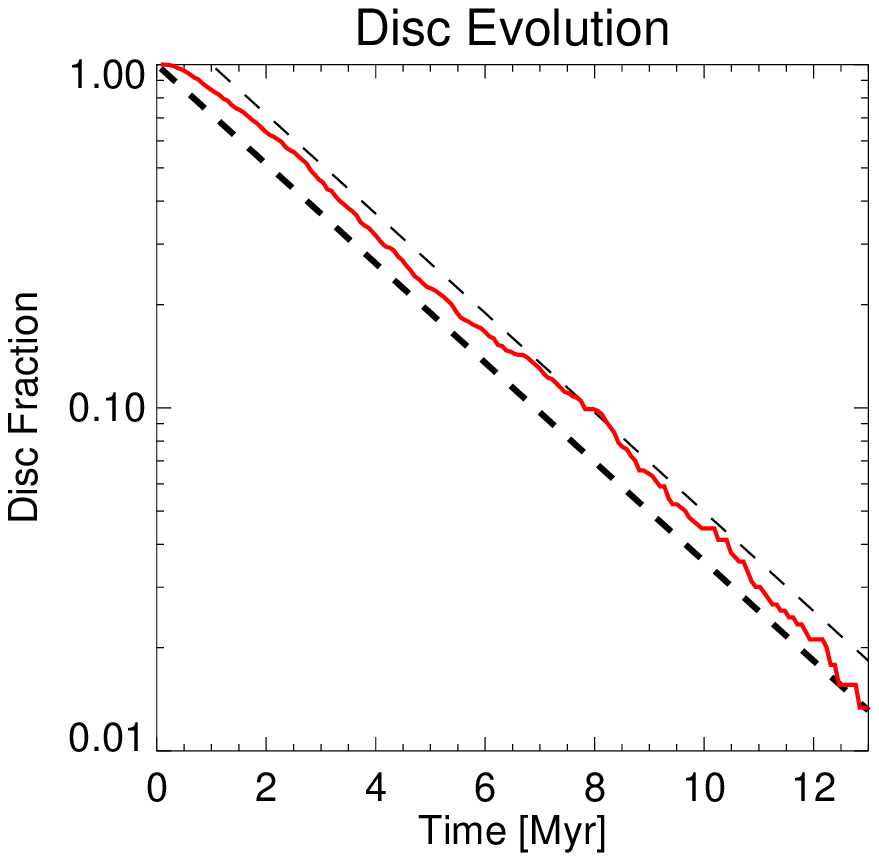}
 \caption{Simulated fraction of stars with discs as a function of time
   (red). The two power-law curves show approximate bounds within which
   the red line should fall to describe observational disc fraction dependence
 on the star's age reasonably well.}
   \label{Sfig_disc}
 \end{figure}

Fig. \ref{Sfig_disc} shows the disc fraction in our model (red) versus time
for a sample of 3000 discs without planets. The black curves show $f_{\rm
  disc0} = \exp(-t/\tau_{\rm disc})$, with $\tau_{\rm disc} = 3
$~Myr. However, the age of a protostar may be uncertain by $\sim 1$~Myr. To
take this uncertainty into account, we also consider $f_{\rm disc1} = 1$ for
$t < 1$~Myr, and $f_{\rm disc1} = \exp(-t'/\tau_{\rm disc})$, where $t' = t -
1$~Myr, to be a reasonable fit to the data.  The simulated disc fraction
dependence with time provides a reasonably close description of the data,
falling in-between the two power-law lines ($f_{\rm disc0}$ and $f_{\rm disc1}$).

\subsection{Planet evolution module}\label{S:planet}

The planet evolution module is nearly identical to that described in section 4
of \cite{Nayakshin15c}, with changes listed in section 2.2.2 of
\cite{NayakshinFletcher15}. In brief, the planet formation module calculates,
in 1D spherically symmetric approximation, the evolution of the internal
structure of a gas fragment of an initial mass $M_0$ with grains treated in a
second-fluid approximation. Since gas accretion onto the fragment is assumed
to be inefficient \citep{NayakshinCha13}, the fragment has a fixed gas mass,
$(1-Z_0)M_0$, where $Z_0$ is the initial metallicity of the fragment and the
surrounding protoplanetary disc. However, the fragment accretes grains by
pebble accretion \citep{LJ14} from the disc at the rate, $\dot M_z$, determined
by the disc evolution module of the code. The pebbles are deposited in the
outermost layers of the fragment. Due to this, the metallicity of the
fragment, $Z$, increases with time. The grains are allowed to grow via grain
sticking collisions and sediment towards the centre
\citep{Boss97,HelledEtal08}. The internal structure of the core in the centre
of the fragment is not modelled, assuming simply a sphere with a constant
material density $\rho_0 = 3$~g~cm$^{-3}$.

The treatment of grains inside the gas fragment follows sections 4.2, 4.3, and
4.6 of \cite{Nayakshin15c}. Three grain species -- water, rocks and CHON --
are followed. CHON is a material made of carbon, hydrogen, oxygen, and
nitrogen, excluding water; the material properties and vapour pressure are
taken to be similar to that of the grains in the coma of Comet Halley
\citep{HelledEtal08}.  The three species are in relative abundances of 0.5,
0.25 and 0.25, respectively.  Grains grow by sticking collisions, although
they fragment if they sediment at too high velocity (see below).  Turbulence
and convective grain mixing are also included, and oppose grain sedimentation.
Grains vaporise once the temperature exceeds the vaporisation temperature for
given species. Core assembly by grain sedimentation therefore depends
sensitively on the temperature and convection in the inner regions of the gas
fragment.  Accretion of grains onto the core is modelled as in section 4.4 of
\cite{Nayakshin15c}.

Since the energy transport within the gas fragment is dominated by convection
soon after its formation \citep{HelledEtal08}, I assume the fragments to be
isentropic. For such a fragment, and a given equation of state (EOS), there
exists a relation between the central temperature, $T_c$, and the total energy
of the fragment, $E_{\rm tot}$. At time $t=0$, an initial value for $T_c$ is
specified. The hydrostatic balance equation for the fragment is solved to
determine the initial $E_{\rm tot}$ and the fragment's radius, $R_p$. The
fragment's luminosity is also determined at this step. $E_{\rm tot}$ is then
evolved according to equation (2), and the procedure repeated again.

The equation of state for the gas includes molecular Hydrogen's vibrational
and rotational degrees of freedom, molecule dissociation, and Hydrogen atom
ionisation at high temperatures. Latent heat of grain vaporisation is also
included.

\subsection{Core growth and luminosity}\label{S:cores}

Core growth via grain sedimentation in my model is indirectly controlled by
two parameters -- the turbulence parameter $\alpha_d$ and the breaking
velocity $v_{\rm br}$. The former sets the diffusive grain mixing by
turbulence \citep{Nayakshin15c}, which opposes grain sedimentation.  In
\cite{NayakshinFletcher15}, the turbulent diffusion parameter $\alpha_d$ was a
Monte Carlo variable with limits from $10^{-4}$ to $10^{-2}$. For larger
values of $\alpha_d$ in this range, turbulence slows down grain assembly
significantly.  In this paper we wish to study the limit of a rapid assembly
of massive solid cores, thus we use a fixed value of $\alpha_d =10^{-4}$. This
is low enough to provide a minimal resistance to grain sedimentation via
turbulent diffusion. Note that this does not affect the convective grain
mixing, which still offers a significant impediment to grain sedimentation,
especially once the core is luminous enough.

The breaking velocity of grains is set by their material
properties. Collisions of grains with velocities $v < v_{\rm br}$ are assumed
sticking, leading to grain growth, whereas collisions with velocities
exceeding $v_{\rm br}$ lead to grain fragmentation and hence decrease in the
grain size. This limits average grain size to a few cm, typically. Breaking
velocity is a Monte-Carlo variable, with upper and lower limits of 15 and 30
m~s$^{-1}$ (cf. Table S1).

Since the internal structure of the core is not explicitly modelled, the
core's luminosity cannot be calculated from first principles. The situation
studied here -- a rapid, e.g., $\sim 10^4$ to $\sim 10^5$ years, assembly of a
core as massive as $10 \mearth$ inside a dense hot gas envelope -- is
significantly different from that envisaged by \cite{Safronov78} in CA
theory. In the classical CA, the gas-free assembly of terrestrial planets was
thought to take as long as $\sim 10^8$ years. In that case it is reasonable to
assume that the accretion energy of grains or planetesimals falling onto the
core is radiated away ``immediately'' \citep{Safronov78}. In the problem at
hand, however, grain accretion energy is first deposited into exciting the
internal degrees of freedom of molecules and atoms, e.g., ionization of upper
loosely bound energy levels of atoms, and the general heating of the core. The
energy will be eventually emitted by the core, of course, as the core cools
and contracts, but a rigorous calculation of the rate of this process is
beyond what is possible at the present.

I follow the approach of \cite{Nayakshin15c,NayakshinFletcher15}, in which
$L_{\rm core}$ represents the delayed grain accretion luminosity of the core.
The core emits its energy of accretion on a finite time scale, $t_{\rm kh}$,
where $t_{\rm kh}$ is free parameter of the model, the Kelvin-Helmholtz
contraction time of the core:
\begin{equation}
L_{\rm core} = {E_{\rm core}\over t_{\rm kh}}\;,
\label{lcore1}
\end{equation}
where $E_{\rm core}$ is the residual potential energy of the core, which is
integrated in time according to
\begin{equation}
{dE_{\rm core} \over dt} = {G M_{\rm core} \dot M_c\over R_{\rm core}} -
{E_{\rm core}\over t_{\rm kh}}\;.
\label{ecore2}
\end{equation}
$M_{\rm core}$ and $R_{\rm core}$ are the running (current) core's mass and
radius. These two equations implement a delayed release of total energy $G
M_{\rm core} /2 R_{\rm core}$ at $t = \infty$, in which case $M_{\rm core}$
and $R_{\rm core}$ are the final mass and radius of the core. In the limit of
$t_{\rm kh}\rightarrow 0$ the energy release described by equation
\ref{ecore2} is given by the instantaneous accretion luminosity, as in the
usual approach in the Core Accretion theory. For simulations ST, $t_{\rm kh} =
3\times 10^5$ years, whereas it is $t_{\rm kh} = 3\times 10^{10}$ years for
runs DC.

\subsection{Population synthesis parameter range}\label{S:popsyn}

To enact a statistically meaningful study, I define a number of Monte-Carlo
variables with a lower and an upper limit (see Table S1).  For each
Monte-Carlo variable, the distribution is uniform in the $\log$ of the
parameter, except for host star metallicity, and is distributed between the
minimum and the maximum values given in Table 1. For example, the initial
protoplanetary disc mass varies between $0.075 \msun$ and $0.2\msun$, while
the disc viscosity parameter is sandwiched between $5\times 10^{-3} < \alpha_0
< 0.05$. 

For the host star metallicity distribution, a Gaussian distribution
is used,
\begin{equation}
{dp(Z_L) \over dZ_L} = {1\over \sigma (2\pi)^{1/2}} \exp\left[ - {Z_L^2\over 2
    \sigma^2}\right]
\label{zdist}
\end{equation}
where $Z_L = $[Fe/H] = [M/H], the usual logarithmically defined metallicity,
$\sigma = 0.22$, and $dp/dZ_L$ is the probability density.  $Z_L \equiv
\log_{10} (Z/Z_\odot)$, where $Z_\odot = 0.015$ is the Solar metallicity, that
is, the fraction of mass in astrophysical metals compared to the total mass of
the Sun. Obviously, $Z = Z_\odot 10^{Z_l} = Z_\odot 10^{\rm [M/H]}$.

As explained in the main text, runs ST, DC and NC are each comprised of 3000
individual planet formation experiments. The run DC (``dim cores'') is
different from ST only in that the Kelvin-Helmholtz time for the core is set to
$t_{\rm kh} = 3 \times 10^{10}$ years, rendering the cores too dim to affect
the host fragment. In the simulation NC (``no cores''), core formation is
artificially suppressed and grain growth disallowed.

\section{Self-disruption of gas fragments at $a > 100$ AU.}\label{S:cold}

As pointed out in the main text, gravitational instability may be effective at
distances well over 100~AU. For example, simulations of \cite{VB06} show
fragment formation on scales of hundreds of AU. Here we show that fragments
born by GI at distances greater than HL Tau rings may also form massive cores
by self-disruption of the young low mass gas clumps. Figure S.2 shows how a
gas fragment initialised at $a=150$~AU evolves in our disc. Since the disc
surface density is much lower at these distances, the fragment migrates in
much slower than the fragments born at $a\simlt 100$~AU do. Pebble accretion
rates are also lower (note the decrease in pebble accretion ``luminosity'',
$L_{\rm peb}$ in panel e of fig. S.2), hence the core growth is slower.

Nevertheless, the end result is similar: the fragment is eventually
disrupted at time $t\approx 0.8$~Myr, when the core mass reaches just above
$5\mearth$.

\begin{figure}
\centering
\includegraphics[width=3.in,angle=0]{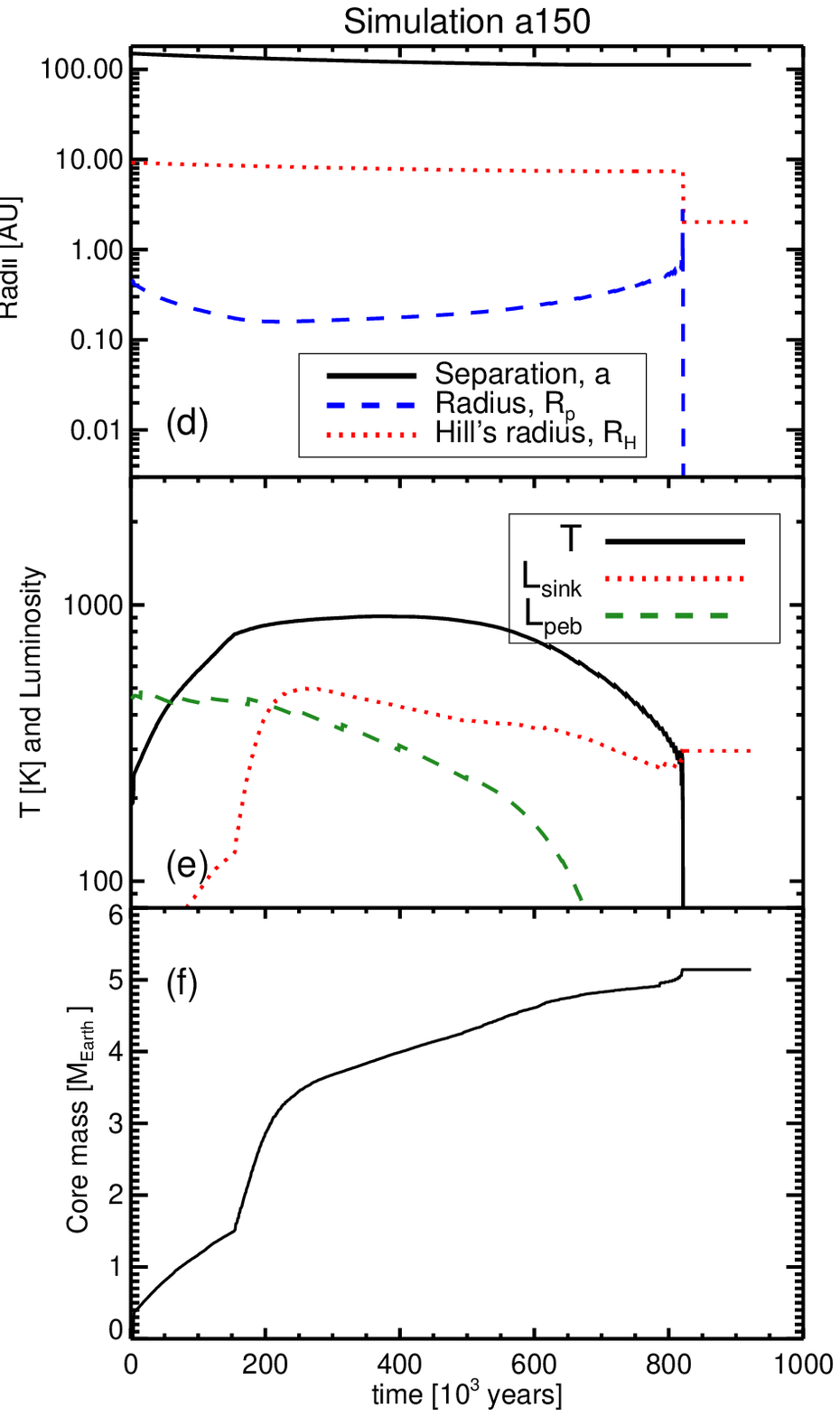}
 \caption{Simulation performed with same physics as the ST series of runs but
   with the fragment started at $a=150$~AU. The fragment self-destructs and
   deposits the core of mass $\approx 5 \mearth$ back into the disc at $a =
   112$~AU, where it remains.}
   \label{Sfig_a150}
 \end{figure}

\begin{table*}
\begin{tabular}{lcccccccccc}
Parameter & $M_0$ & $\zeta_{\rm ev}$ &   $f_{\rm p}$ & $a_0$  & $M_{\rm d}$ &
$f_{\rm migr}$ & $t_{\rm kh}$ & $\alpha_{d}$ & $v_{\rm br}$ & $\alpha_0$ \\ \hline
Min &  $1/3$ & 0.02 &  0.04 &  70  &  0.075 & 1  &  $3 \times 10^5$ &
$10^{-4}$ & 15 & $5\times 10^{-3}$ \\
Max &  8& 3.0  &  0.08  &  105 &  0.2  & 4 &  $3 \times 10^5$   & 10$^{-4}$ &
30 & $5\times 10^{-2}$\\ \hline
\end{tabular}
\label{tab:1}
\caption{The range of the Monte Carlo parameters of the ``standard''
  population synthesis calculation (run ST) for this paper. The first row
  gives parameter names, the next two their minimum and maximum values. The
  columns are: planet's initial mass, $M_0$, in Jupiter masses; $\zeta_{\rm
    ev}$, the evaporation rate factor; $f_{\rm p}$, the pebble mass fraction
  determining the fraction of the disc grain mass in pebbles; $a_0$ [AU], the
  initial position of the fragment; $M_{\rm d}$, the initial mass of the disc,
  in units of $\msun$; $f_{\rm migr}$, the type I planet migration factor;
  $t_{\rm kh}$, in years, determines the luminosity of the core; $\alpha_d$,
  turbulence parameter within the fragment; $v_{\rm br}$, the grain breaking
  velocity, in m/s; the disc viscosity parameter, $\alpha_0$.}
\end{table*}

\bsp	
\label{lastpage}
\end{document}